\documentclass[aps, prd, twocolumn, showpacs, showkeys, preprintnumbers, bibnotes, floatfix, longbibliography, notitlepage, nofootinbib]{revtex4-2}
\pdfoutput=1
\usepackage[colorlinks=true,breaklinks=true]{hyperref}
\usepackage{amsmath}
\usepackage{amsfonts}
\usepackage{amssymb}
\usepackage{graphicx}
\usepackage[T1,OT1]{fontenc} 
\usepackage[utf8]{inputenc}
\DeclareTextCommandDefault{\k}[1]{{\fontencoding{T1}\selectfont\k{#1}}}
\usepackage{booktabs}
\usepackage{multirow}
\usepackage{enumitem}
\usepackage[dvipsnames]{xcolor}
\usepackage{slashed}
\definecolor{darkblue}{rgb}{0,0,0.5}
\definecolor{firebrick}{rgb}{0.75,0.125,0.125}
\definecolor{darkgreen}{rgb}{0,0.5,0}
\hypersetup{urlcolor=darkblue, citecolor=darkgreen, linkcolor=firebrick}
\usepackage[capitalise]{cleveref}
\usepackage{orcidlink}

\newcommand{\ie}{{\it i.e.}}
\newcommand{\eg}{{\it e.g.}}

\newcommand{\eq}{Eq.}

\newcommand{\fig}{Fig.}
\newcommand{\Refe}{Ref.}
\newcommand{\Refes}{Refs.}
\newcommand{\equ}[1]{\eq~(\ref{equ:#1})}
\newcommand{\figu}[1]{\fig~\ref{fig:#1}}
\newcommand{\Uhat}{\hat U}
\newcommand{\meff}{m_{\rm eff}}

\graphicspath{{figures/}}

\begin{document}

\title{The renormalization-group running of neutrino mixing is undetectable}

\makeatletter\def\@listand{}\makeatother 
\author{Gabriela Barenboim\,\orcidlink{0000-0002-3249-7467}}
\email{gabriela.barenboim@uv.es}
\affiliation{Departament de F\'isica Te\`orica and IFIC, Universitat de Val\`encia-CSIC, E-46100, Burjassot, Spain}

\author{Mauricio Bustamante\,\orcidlink{0000-0001-6923-0865}}
\email{mbustamante@nbi.ku.dk}
\affiliation{Niels Bohr International Academy, Niels Bohr Institute,\\University of Copenhagen, 2100 Copenhagen, Denmark}

\author{Qinrui Liu\,\orcidlink{0000-0003-3379-6423}}
\email{qinrui\_liu@sfu.ca}
\affiliation{Department of Physics, Simon Fraser University, Burnaby, BC V5A 1S6, Canada}
\affiliation{Arthur B.~McDonald Canadian Astroparticle Physics Research Institute, Kingston, ON K7L 3N6, Canada}

\date{\today}

\begin{abstract}
Quantum corrections make the parameters of a theory depend on an arbitrary renormalization scale, $\mu$. The renormalization-group (RG) running of neutrino masses and mixing is a standard tool to relate models defined at a high mass scale to measured values. A growing body of work holds that the running is also directly observable: a neutrino produced or detected at momentum transfer $Q$ would carry the mixing matrix evaluated at $\mu = Q$. Proposed signatures include flavor change at zero distance from the source, shifts in the measured CP-violation phase, and an anomalous flavor composition of high-energy astrophysical neutrinos. We show that this premise is incorrect. The mixing is a property of the neutrino itself, fixed by its self-energy on the mass shell, not by the process that produces or detects it. Setting $\mu = Q$ is legitimate, but then large logarithms in the self-energy undo the running; the premise incorrectly drops them. Hence, apart from small vertex and box corrections, every experiment measures the same mixing: the running is undetectable. We illustrate this in the Minimal Supersymmetric Standard Model, two seesaw models, and the Standard Model effective field theory. Independently, short-baseline data already limit the change of any one mixing angle between pion decay and detection at $Q$ of a few GeV to below $0.7^\circ$--$2.4^\circ$.
\end{abstract}

\maketitle

\section{Introduction}
\label{sec:intro}

\begin{figure*}[t!]
 \centering
 \includegraphics[width=\textwidth]{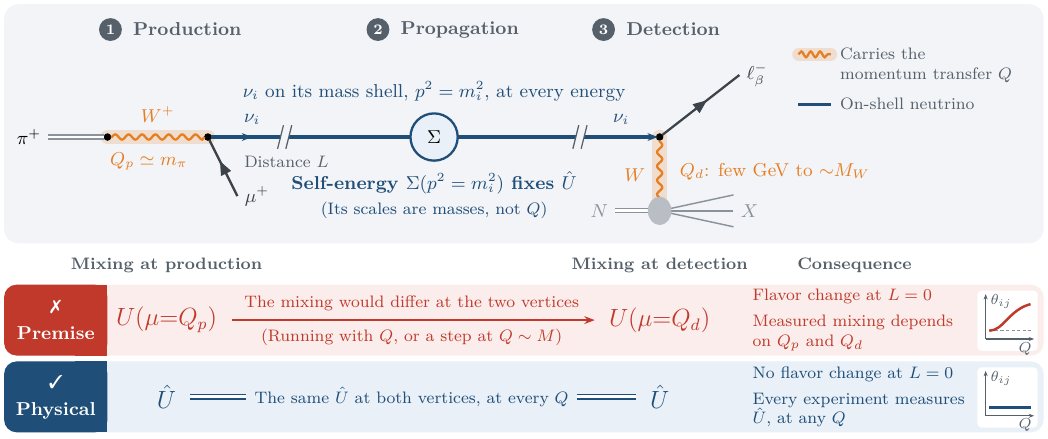}
 \caption{\textbf{\textit{Which mixing matrix enters neutrino production and detection.}} \textit{Top:} an example process. A neutrino, $\nu_i$, is produced in pion decay, at momentum transfer $Q_p \simeq m_\pi$, and travels a macroscopic distance, $L$. It is detected in deep-inelastic scattering, $\nu_i + N \to \ell_\beta + X$, at momentum transfer $Q_d$, from a few GeV in fixed-target experiments up to about the $W$ mass, $M_W \approx 80$~GeV, in IceCube. The momentum transfer is set by the four-momentum of the $W$ bosons; the neutrino stays on its mass shell, $p^2 = m_i^2$, at every energy and momentum transfer. The neutrino propagator, including its self-energy, $\Sigma$, at $p^2 = m_i^2$, fixes the physical mixing matrix, $\Uhat$. The self-energy depends on the masses of the particles in its loop, not on the momentum transfer. The same holds for any production or detection process. \textit{Bottom:} the works in Table~\ref{tab:literature} instead take the mixing at production and detection to be the running mixing matrix at $\mu = Q_p$ and $\mu = Q_d$, or to change in a step when $Q$ crosses the mass $M$ of a heavy neutral lepton. This premise predicts flavor change at zero distance from the source and a measured mixing that depends on the momentum transfer. Apart from flavor-universal corrections, however, the momentum transfer enters only the vertex and box corrections, which are small and lack the amplification by nearly degenerate masses that makes the running large (Sec.~\ref{sec:general}). Hence, every experiment, at any momentum transfer, measures the same $\Uhat$.}
 \label{fig:schematic}
\end{figure*}

Neutrino mixing is the mismatch between the neutrino states that participate in weak interactions---the flavor states, $\nu_e$, $\nu_\mu$, and $\nu_\tau$---and the states that propagate---the mass eigenstates, $\nu_1$, $\nu_2$, and $\nu_3$. Each flavor state is a superposition of the mass eigenstates, $|\nu_\alpha\rangle = \sum_i U_{\alpha i}^\ast |\nu_i\rangle$, where $U$ is the lepton mixing matrix, or Pontecorvo--Maki--Nakagawa--Sakata (PMNS) matrix~\cite{Pontecorvo:1957cp, Maki:1962mu}. The mixing matrix is described by the mixing parameters: three mixing angles, $\theta_{12}$, $\theta_{23}$, and $\theta_{13}$, a CP-violation phase, $\delta_{\rm CP}$, and, if neutrinos are their own antiparticles, \ie, Majorana particles, two Majorana phases. Because the mass eigenstates have different masses, they acquire different phases as they propagate. As a result, a neutrino produced with one flavor can be detected with another, with a probability that oscillates with the distance traveled, and that depends on the mixing angles, $\delta_{\rm CP}$, and the differences between the squared neutrino masses.

The values of the mixing parameters presumably carry information about the origin of neutrino masses, the flavor structure of leptons, and the matter-antimatter asymmetry of the Universe~\cite{Moffat:2018smo}. Models of neutrino mass predict these values, which are contrasted against experimental measurements. Today, a combination of solar, reactor, accelerator, and atmospheric neutrinos, from MeV to TeV energies, constrains the mixing angles and $\delta_{\rm CP}$ to within a few to tens of percent~\cite{deSalas:2020pgw, Capozzi:2021fjo, Esteban:2024eli}. High-energy astrophysical neutrinos extend tests of mixing to PeV energies~\cite{AlvesBatista:2021eeu, Ackermann:2022rqc}, although they do not yet constrain the mixing parameters~\cite{IceCube:2025ole, Bustamante:2026aur}.

In the Lagrangian of a model of neutrino mass, the coefficients of the neutrino mass terms form a matrix, the neutrino mass matrix. Diagonalizing this matrix yields the masses of the mass eigenstates, $m_1$, $m_2$, and $m_3$, and the mixing matrix. Quantum corrections, \ie, loop contributions to a process, contain integrals that diverge at large loop momenta. Renormalization absorbs the divergences into the Lagrangian parameters, including the mass matrix. In the widely used modified minimal-subtraction ($\overline{\rm MS}$) scheme~\cite{tHooft:1973mfk, Weinberg:1973xwm, Bardeen:1978yd}, the parameters are defined at an arbitrary scale, the renormalization scale, $\mu$. Thus, the parameters acquire a dependence on $\mu$, known as renormalization-group (RG) running~\cite{Stueckelberg:1952hma, Gell-Mann:1954yli, Callan:1970yg, Symanzik:1970rt, Wilson:1971bg}.

Observables, \ie, measurable quantities, such as decay rates, cross sections, and oscillation probabilities, are computed, including quantum corrections, from the renormalized parameters. Since $\mu$ is an arbitrary choice, observables cannot depend on it: when $\mu$ changes, the running of the parameters cancels the change in the quantum corrections. Masses illustrate the difference between parameters and observables. The pole mass of a freely propagating particle, \ie, the position of the pole of its propagator, including quantum corrections, is an observable: it depends neither on $\mu$ nor on the renormalization scheme. The $\overline{\rm MS}$ mass of the particle is a parameter: it depends on $\mu$. The two differ by quantum corrections.

The same holds for neutrinos. The masses and mixing obtained by diagonalizing the $\overline{\rm MS}$ neutrino mass matrix are parameters: they run. The mass-squared differences and mixing parameters extracted from oscillation data are observables. However, models of neutrino mass typically predict the values of the parameters at a high scale, \ie, at a value of $\mu$ far above the electroweak scale of $\sim 100$~GeV. Examples are grand unified theories, models with flavor symmetries, and seesaw models~\cite{Minkowski:1977sc, Yanagida:1979as, Gell-Mann:1979vob, Mohapatra:1979ia, Glashow:1979nm, Schechter:1980gr}. To test these predictions, the standard procedure is to run the parameters from the high scale down to the electroweak scale~\cite{Babu:1993qv, Chankowski:1993tx, Ellis:1999my, Haba:1999ca, Casas:1999ac, Casas:1999tg, Balaji:2000gd, Antusch:2001ck, Antusch:2001vn, Chankowski:2001mx, Antusch:2002rr, Antusch:2003kp, Antusch:2005gp, Ohlsson:2013xva}. For nearly degenerate neutrino masses, the running of the mixing angles can be large~\cite{Ellis:1999my, Casas:1999ac, Balaji:2000gd, Antusch:2003kp}. In the Standard Model with massive neutrinos, the running becomes negligible below the electroweak scale: below their masses, the $W$, $Z$, and Higgs bosons, whose loops drive it, no longer contribute. There, the parameters reproduce the observables extracted from oscillation data, up to small quantum corrections.

\textbf{A different idea has recently gained ground: that the running is directly observable.} In this view, the mixing that enters the production or detection of a neutrino in a charged-current interaction, \ie, one mediated by a $W$ boson, depends on the momentum transfer of the interaction, $Q$. Here, $Q^2 = |q^2|$, where $q$ is the four-momentum carried by the $W$, so that $Q^2 = -q^2$ in neutrino scattering and $Q = m_\pi$, the pion mass, in neutrino production via pion decay. The dependence is usually introduced by evaluating the running mixing matrix at $\mu = Q$. This choice follows standard practice for running couplings, such as the strong coupling, $\alpha_s$, for which setting $\mu = Q$ avoids large logarithms of $Q/\mu$ in the quantum corrections to a process. Since production and detection of a neutrino often occur at different $Q$, the two ends of an oscillation experiment would use different mixing matrices. \textbf{We call this idea the premise.} Since 2011, a number of works have adopted it (Sec.~\ref{sec:literature}).

Figure~\ref{fig:schematic} shows the main result of this paper: \textbf{the premise is incorrect.} The mixing that enters the production and detection of a neutrino is not the mixing matrix of the Lagrangian evaluated at $\mu = Q$. Instead, it is fixed by the neutrino propagator, including its quantum corrections, \ie, the neutrino self-energy. The neutrino travels a macroscopic distance between production and detection. Therefore, each of its mass eigenstates, $\nu_i$, is on its mass shell: its four-momentum, $p$, satisfies $p^2 = m_i^2$, whatever its energy is. The self-energy depends on the four-momentum of the neutrino, $p$, but not on that of the $W$, $q$, which sets $Q$. Momentum conservation relates $p$ to $q$, but it leaves $p^2 = m_i^2$ unchanged. Hence, at $p^2 = m_i^2$, the self-energy depends on the masses of the particles in the loop, but not on $Q$. \textbf{Therefore, the mixing is a property of the neutrino, not of the process that produces or detects it.} Production and detection share a single mixing matrix, which we call the physical mixing, $\Uhat$.

This does not contradict the RG running of the mixing. The running concerns $U(\mu)$, the mixing matrix as a parameter of the Lagrangian. The renormalization scale, $\mu$, is only a bookkeeping device of the calculation: it can take any value, including $Q$, without changing any observable. When $\mu$ changes, both $U(\mu)$ and the self-energy change, in opposite ways; their combination, the physical mixing, $\Uhat$, does not change. In particular, with $\mu = Q$, as in the premise, the production and detection amplitudes contain $U(Q)$, but the self-energy then acquires large logarithms of the ratio of $Q$ to the masses of the particles in the loop. These logarithms rotate $U(Q)$ back to $\Uhat$. The premise is incorrect because it keeps $U(Q)$ but drops these counteracting logarithms (Sec.~\ref{sec:general-muQ}). \textbf{Hence, the RG running of the mixing is undetectable.}

Once the logarithms in the self-energy undo the running in $U(Q)$, the only dependence on $Q$ that can affect the flavor comes from the vertex and box diagrams at production and detection. These corrections are small, suppressed by loop factors and couplings. They also lack the amplification that makes the running of the mixing large. That amplification occurs only in the self-energy, where nearly degenerate masses turn small corrections to the mass matrix into large rotations of the mixing.

Section~\ref{sec:literature} describes the works that adopt the premise and how they justify it. Section~\ref{sec:synopsis} breaks the argument into steps, with an analogy. Section~\ref{sec:general} derives the result in general. Section~\ref{sec:antusch} compares it with the standard treatment of RG running and shows what the running is good for. Section~\ref{sec:examples} works out four examples in detail. Section~\ref{sec:data} shows what existing data already say. Section~\ref{sec:misconceptions} addresses common misconceptions. Section~\ref{sec:legit} discusses what can legitimately make the measured flavor differ from expectations. Section~\ref{sec:conclusions} concludes. The appendices contain technical details.

\section{The premise in the literature}
\label{sec:literature}

\begin{table*}[t]
 \footnotesize
 \caption{\label{tab:literature}\textbf{\textit{Works that take the mixing in production or detection to depend on the momentum transfer, $Q$, of the process.}} In all but one of them, the mixing at production and at detection is the running mixing matrix evaluated at the respective momentum transfer, $U(\mu = Q)$. In \Refe~\cite{Barenboim:2026mvn}, the mixing changes in a step at $Q = M$, the mass of a heavy neutral lepton (HNL); this is the discrete counterpart of the same assumption (Sec.~\ref{sec:examples-hnl}). The signatures listed do not arise from RG running, nor from threshold corrections that switch on or off with $Q$ (Secs.~\ref{sec:general}--\ref{sec:data}). Abbreviations: SM, Standard Model; MSSM, Minimal Supersymmetric SM; SMEFT, SM effective field theory; $\tan\beta$, ratio of the vacuum expectation values of the two Higgs doublets of the MSSM; $N$, sterile neutrino; scotogenic, with neutrino masses generated at one loop; ND, near detector, \ie, a detector close to the source; TAO, Taishan Antineutrino Observatory, the near detector of JUNO; FASER$\nu$2, a proposed neutrino detector at the Large Hadron Collider.}
 \begin{ruledtabular}
  \begin{tabular}{p{0.36\textwidth} p{0.24\textwidth} p{0.34\textwidth}}
   Reference & Setting & Proposed signature \\
   \hline
   Bustamante, Gago, Jones P\'erez (2011)~\cite{Bustamante:2010bf} & MSSM, large $\tan\beta$ & Flavor composition of high-energy astrophysical neutrinos \\
   Babu, Brdar, de Gouv\^ea, Machado (2022)~\cite{Babu:2021cxe} & Light $N$ and scalar (scotogenic and seesaw variants) & T2K, NOvA; $\theta_{13}$ and $\theta_{23}$ mismatches between experiments; zero-baseline transitions; IceCube flavor composition \\
   Babu, Brdar, de Gouv\^ea, Machado (2023)~\cite{Babu:2022non} & eV-scale sterile neutrinos with a light $U(1)^\prime$ boson & LSND and MiniBooNE appearance anomalies \\
   Ge, Kong, Pasquini (2024)~\cite{Ge:2023azz} & Model-independent running & CP-phase measurement at DUNE with mismatched $Q_p$, $Q_d$; zero-baseline transitions \\
   Ge, Kong, Pasquini (2025)~\cite{Ge:2024ibn} & Model-independent running & Zero-baseline $\bar\nu_e$ deficit from running $\delta_{\rm CP}$ at a reactor near detector \\
   Mir, Arg\"uelles, Babu, Brdar (2026)~\cite{Mir:2025fae} & Light new particles coupled to neutrinos & Excess of $\nu_\tau$ double bangs in IceCube \\
   Bustamante, Liu, Barenboim (2026)~\cite{Bustamante:2026zst} & SM, MSSM, SMEFT & Flavor composition of high-energy astrophysical neutrinos \\
   Barenboim, Bustamante, Liu (2026)~\cite{Barenboim:2026mvn} & GeV-scale Majorana HNL; threshold at $Q = M$ & Flavor composition of high-energy astrophysical neutrinos \\
   Ge, Pinheiro, Qin (2026)~\cite{Ge:2026hsh} & Model-independent running & Mass-ordering sensitivity and running of Majorana phases at JUNO, JUNO-TAO \\
   Denton, Ge, Kong, Pasquini (2026)~\cite{Denton:2026mkp} & Model-independent running & Zero-baseline transitions at DUNE-ND, JUNO-TAO, FASER$\nu$2 \\
  \end{tabular}
 \end{ruledtabular}
\end{table*}

Table~\ref{tab:literature} lists the works in which we identified the premise. To our knowledge, the premise first appeared in a study of the Minimal Supersymmetric Standard Model (MSSM) with high-energy astrophysical neutrinos~\cite{Bustamante:2010bf}. It has since been applied to models in which new light states make the running large below the electroweak scale~\cite{Babu:2021cxe}; to mixing with sterile neutrinos, which do not take part in weak interactions, to explain anomalies in short-baseline experiments~\cite{Babu:2022non}; to measurements of $\delta_{\rm CP}$ at the Deep Underground Neutrino Experiment (DUNE)~\cite{Ge:2023azz}; to tests of the running of $\delta_{\rm CP}$ with reactor neutrinos~\cite{Ge:2024ibn}; to the combination of short-baseline experiments at different energies to constrain the running of mixing parameters~\cite{Denton:2026mkp}; to the Jiangmen Underground Neutrino Observatory (JUNO), including its sensitivity to the neutrino mass ordering and a claimed hint in its data that neutrinos are Majorana particles~\cite{Ge:2026hsh}; to $\nu_\tau$-induced double-bang events~\cite{Learned:1994wg, MammenAbraham:2022xoc} in IceCube~\cite{Mir:2025fae}; to the flavor composition of TeV--EeV astrophysical neutrinos~\cite{Bustamante:2026zst}; and to corrections to the neutrino mass matrix from heavy neutral leptons (HNLs), with a step in the mixing when $Q$ crosses the HNL mass~\cite{Barenboim:2026mvn}.

Five of the works in Table~\ref{tab:literature} that evaluate the running mixing matrix at $\mu = Q$ justify this choice explicitly, in three ways:
\begin{enumerate}[leftmargin=*, label=(\arabic*)]
 \item
  Vertex corrections are evaluated with $\mu \approx Q$ to avoid large logarithms; all parameters in the vertex must then be evaluated at the same scale, including the mixing. This is the argument of \Refe~\cite{Bustamante:2010bf}, which describes the running as ``an independent second source of corrections to the vertex.''

 \item
  Reference~\cite{Babu:2021cxe} assumes, following the RG approach, that one can replace $g U_{\alpha i}$ by $g(Q^2) U_{\alpha i}(Q^2)$ in the coupling of the $W$ boson to a charged lepton of flavor $\alpha$ and $\nu_i$. Here, $g$ is the weak coupling constant and $U_{\alpha i}$ is an element of the mixing matrix. It argues that the propagating neutrino is on shell, with pole masses, so that production and detection can be treated separately. Reference~\cite{Ge:2023azz} adds that, since the renormalization scale of the on-shell propagator is fixed at the neutrino masses, the running of the mixing should come from the charged-current vertex.

 \item
  References~\cite{Ge:2023azz, Ge:2024ibn, Denton:2026mkp} identify the choice $\mu = Q$ with the Gell-Mann--Low scheme~\cite{Gell-Mann:1954yli}. In this scheme, a coupling is defined by an observable at momentum transfer $Q$, so that its dependence on $Q$ is physical.
\end{enumerate}
The other works that use the running adopt the premise from these~\cite{Babu:2022non, Mir:2025fae, Ge:2026hsh}, or appeal to the running of couplings in general~\cite{Bustamante:2026zst}. 

Reference~\cite{Barenboim:2026mvn} distinguishes its effect from RG running: in it, the mixing changes in a step when $Q$ crosses the HNL mass, $M$. It argues that detection at $Q > M$ resolves the HNL, so that the mixing above $Q = M$ differs from that below it by the effect of the HNL correction to the mass matrix. Treating the scale at which the HNL is integrated out, $\mu \simeq M$, as the momentum transfer of the process is the premise in discrete form (Sec.~\ref{sec:examples-hnl}).

None of these works computes the vertex or box corrections to the charged-current amplitude. These are the only corrections that can make its flavor structure depend on $Q$ (Sec.~\ref{sec:general-amplitude}). Reference~\cite{Bustamante:2010bf} notes that, in the MSSM, such corrections could change the flavor structure of the vertex, but it evaluates only the effect of the running. In nine of the works, the $Q$-dependence comes instead from the running of the mixing parameters, evaluated at $\mu = Q$. Five of them obtain the running by solving RG equations~\cite{Bustamante:2010bf, Babu:2021cxe, Babu:2022non, Mir:2025fae, Bustamante:2026zst}; four describe it by $\beta$ functions---which drive the RG running---treated as free constants~\cite{Ge:2023azz, Ge:2024ibn, Ge:2026hsh, Denton:2026mkp}. In \Refe~\cite{Barenboim:2026mvn}, the size of the step is set by a finite one-loop correction to the light-neutrino mass matrix, generated by a neutrino self-energy.

\section{The argument, step by step}
\label{sec:synopsis}

Section~\ref{sec:intro} gave the argument against the premise in brief. The mixing that enters production and detection is fixed by the neutrino propagator, including its self-energy, evaluated on the mass shell, $p^2 = m_i^2$. Hence, the mixing does not depend on the momentum transfer, $Q$. Setting $\mu = Q$ does not change the conclusion: the self-energy then contains large logarithms, which cannot be neglected: they rotate $U(Q)$ back to the physical mixing, $\Uhat$. The only dependence on $Q$ that can affect the flavor comes from small vertex and box corrections at production and at detection. Below, we illustrate the argument with an analogy, separate four scales that are easy to conflate, and break the argument into six steps, each derived in a later section.

\smallskip

\textit{An analogy.---}A neutral kaon that travels a macroscopic distance is a $K_L$ or a $K_S$. The $K^0$--$\bar K^0$ composition of each of them follows from the kaon propagator, including the self-energy of the kaon, generated in part by box diagrams with $W$ bosons and heavy quarks~\cite{Gaillard:1974hs, Inami:1980fz}. This composition does not depend on the momentum transfer of the collision in which the kaon is later detected: the kaon arrives as a $K_L$ or a $K_S$, whatever the collision. Neutrino mass eigenstates behave in the same way. The flavor content of each of them follows from the neutrino propagator, including its self-energy. Since the mixing describes the composition of the incoming neutrino, not a coupling of the vertex, the premise amounts to assuming that this composition depends on the momentum transfer of the collision that detects the neutrino.

\smallskip

\textit{Four scales.---}Table~\ref{tab:scales} shows four different quantities whose conflation causes much of the confusion: the renormalization scale, $\mu$; the momentum transfer of the production or detection process, $Q$; the invariant mass of the propagating neutrino, $\sqrt{p^2} = m_i$; and the neutrino energy, $E$. The premise identifies $\mu$ with $Q$. Since $Q$ typically grows with $E$ in deep-inelastic scattering, the resulting dependence is often described as a dependence on energy~\cite{Bustamante:2010bf, Babu:2021cxe, Babu:2022non}. Arguments that detection at large $Q$ resolves heavy states~\cite{Barenboim:2026mvn} treat $Q$ as a property of the neutrino line; in fact, $Q$ is set by the four-momentum of the $W$, while the neutrino stays at $p^2 = m_i^2$ (Sec.~\ref{sec:misconceptions}). Of the four quantities, only $\mu$ and $p^2$ enter the neutrino self-energy. The dependence of the self-energy on $\mu$ cancels against that of $U(\mu)$, so the mixing is set by $p^2 = m_i^2$, which is the same in every experiment.

\begin{table*}[t]
 \footnotesize
 \caption{\label{tab:scales}\textbf{\textit{Four scales that are easy to conflate.}} Only the invariant mass of the propagating neutrino sets the mixing that production and detection see.}
 \begin{ruledtabular}
  \begin{tabular}{p{0.05\textwidth} p{0.33\textwidth} p{0.26\textwidth} p{0.30\textwidth}}
   & What it is & Typical values & Does it set the mixing? \\
   \hline
   $\mu$ & Renormalization scale: a bookkeeping choice in a calculation & Any & No; observables do not depend on it \\
   $Q$ & Momentum transfer of production or detection; set by the four-momentum of the $W$ & $m_\pi$ (pion decay); GeV to $M_W$ (deep-inelastic scattering) & No; it enters only corrections that are flavor-universal or small \\
   $p^2$ & Invariant mass squared of the propagating neutrino, $m_i^2$ & $(0.01$--$0.05~{\rm eV})^2$ & Yes; the mixing is defined at $p^2 = m_i^2$ \\
   $E$ & Neutrino energy; depends on the reference frame & MeV--PeV & No; the mixing does not depend on the reference frame \\
  \end{tabular}
 \end{ruledtabular}
\end{table*}

\smallskip

\textit{The argument.---}In six statements:
\begin{enumerate}[leftmargin=*]
  \item
  The columns of the physical mixing matrix, $\Uhat$, are the eigenvectors of the physical neutrino mass matrix. This matrix includes all quantum corrections to the neutrino propagator, evaluated for a neutrino on its mass shell, $p^2 = m_i^2 \simeq 0$. The eigenvalues of this matrix are the pole masses, whose mass-squared differences, $\Delta m_{ij}^2 \equiv m_i^2 - m_j^2$, oscillation experiments measure (Sec.~\ref{sec:general-physical}).

 \item
  The quantum corrections to a charged-current process fall into two groups. The first group contains the corrections to the propagators of the external particles, \ie, of the neutrino and the charged lepton that enter or leave the process. Each depends on the four-momentum of its own particle only through $p^2 = m_i^2$ or $m_\ell^2$, on the masses of the particles in its loop, and on $\mu$. The second group contains the corrections that involve the four-momentum of the $W$, $q$, which sets $Q$: the self-energies of the $W$, which are the same for every flavor, and the corrections at the interaction point, \ie, vertex and box diagrams. Therefore, only the second group depends on $Q$ (Sec.~\ref{sec:general-amplitude}).

 \item
  The corrections to the neutrino propagator in the first group yield the physical mixing of statement~1. Since the first group does not depend on $Q$, this mixing is the same at every $Q$. Therefore, up to the small corrections of the second group, production and detection see the same mixing, $\Uhat$ (Sec.~\ref{sec:general-amplitude}).

 \item
  The second group of corrections cannot mimic running. The self-energies of the $W$ are the same for every flavor, so they cannot rotate the mixing. Large rotations of the mixing come from factors $1/(m_i^2 - m_j^2)$, which are large when two masses are nearly equal. These factors appear only on the external neutrino line, where $p^2 = m_i^2$ lies close to $m_j^2$. Inside vertex or box diagrams, the neutrino momentum is far from the mass shell; there, these factors do not appear. Hence, the corrections at the interaction point lack this enhancement. They are also suppressed by loop factors and couplings, so they do not rotate the mixing appreciably (Sec.~\ref{sec:general-vertex}).

 \item
  The running of the Lagrangian parameter $U(\mu)$ does not contradict statements~1--4. When $\mu$ changes, $U(\mu)$ and the corrections to the neutrino propagator change in opposite ways; their combination, the physical mixing, $\Uhat$, stays fixed. Choosing $\mu = Q$ sets the parameter to $U(Q)$. The corrections to the neutrino propagator then contain large logarithms of $Q/M_X$, where $M_X$ are the masses of the loop particles. Their dependence on $Q$ comes only from the choice $\mu = Q$. These logarithms rotate $U(Q)$ back to $\Uhat$ (Sec.~\ref{sec:general-muQ}).

 \item
  If the premise were true, and if the mixing ran appreciably between the pion mass and a few GeV, short-baseline experiments would have observed flavor change at the source. They did not. Their null results limit the change of any one mixing angle between pion decay and detection at momentum transfers of a few GeV to less than $0.7^\circ$--$2.4^\circ$, depending on the angle. Some effects considered in the literature are far larger: benchmark BP2 of \Refe~\cite{Babu:2021cxe} halves $\theta_{12}$ between the pion mass and 2.5~GeV, which that work notes the same data exclude; for the HNL benchmarks of \Refe~\cite{Barenboim:2026mvn}, the premise predicts that $\theta_{12}$ jumps from $33.8^\circ$ to between $15^\circ$ and $50^\circ$ (Sec.~\ref{sec:data}).
\end{enumerate}
The argument concerns the mixing at production and detection. Effects that act during propagation, such as matter effects, are separate, as are new interactions at production or detection. Section~\ref{sec:legit} discusses them.

\section{The mixing in the charged-current amplitude}
\label{sec:general}

This section derives statements~1--5 of Sec.~\ref{sec:synopsis}. Readers who prefer to skip the formalism can go directly to Sec.~\ref{sec:antusch}.

\subsection{Running parameters}
\label{sec:general-running}

Below the scale of the new physics responsible for neutrino mass, the neutrino mass matrix arises from the Weinberg operator~\cite{Weinberg:1979sa}, the dimension-five operator built from two lepton doublets and two Higgs doublets, with flavor-space coefficient $\kappa(\mu)$. After electroweak symmetry breaking, it yields the Majorana mass term $-\frac{1}{2} \overline{\nu^c_{L}} \, m \, \nu_L + {\rm h.c.}$, with $m(\mu) \propto \kappa(\mu)$. We write the mixing matrix that diagonalizes $m(\mu)$ as $U(\mu)$, \ie, $U^T(\mu) m(\mu) U(\mu) = {\rm diag}(m_1, m_2, m_3)$. We call $U(\mu)$ the running mixing matrix, and the angles and phases that parametrize it, such as $\theta_{ij}(\mu)$, the running mixing parameters. They are parameters of the Lagrangian, not observables. In the basis of the eigenstates of $m(\mu)$, the charged-current interaction is
\begin{equation}
 \label{equ:cc}
 -\mathcal{L}_{\rm CC}
 =
 \frac{g}{\sqrt{2}} \, \overline{\ell_\alpha} \gamma^\mu P_L \, U_{\alpha i}(\mu) \, \nu_i \, W_\mu^- + {\rm h.c.} \;.
\end{equation}
Here, $g$ is the weak coupling constant, $\ell_\alpha$ are the charged leptons, $\nu_i$ are the eigenstates of $m(\mu)$, and $P_L = (1 - \gamma_5)/2$ is the left-chirality projector. At a generic $\mu$, the $\nu_i$ differ from the propagating mass eigenstates of Sec.~\ref{sec:intro}, which we denote by $\hat\nu_i$ (Sec.~\ref{sec:general-physical}).

At one loop, the coefficient of the Weinberg operator obeys~\cite{Babu:1993qv, Chankowski:1993tx, Antusch:2001ck, Antusch:2001vn}
\begin{equation}
 \label{equ:rge}
 16\pi^2 \frac{d\kappa}{dt}
 =
 C \left[ (Y_e^\dagger Y_e)^T \kappa + \kappa \, Y_e^\dagger Y_e \right] + A \kappa \;,
\end{equation}
where $t \equiv \ln(\mu/\mu_0)$, $\mu_0$ is the reference scale at which the initial value of $\kappa$ is set, and $Y_e$ is the charged-lepton Yukawa matrix. The coefficient $C$ depends on the Higgs sector: it is $-3/2$ in the Standard Model (SM) and $1$ in the MSSM, which we use in Sec.~\ref{sec:examples-mssm}. The coefficient $A$ contains the gauge and Higgs couplings and is the same for all flavors. Only the first term changes the mixing; the second rescales $\kappa$ as a whole, leaving its eigenvectors unchanged. The effect of the first term on the running solar mixing angle, $\theta_{12}(\mu)$, is, to leading order~\cite{Casas:1999tg, Antusch:2003kp},
\begin{equation}
 \label{equ:theta12_running}
 \frac{d\theta_{12}}{dt}
 \simeq
 -\frac{C y_\tau^2}{32\pi^2} \sin 2\theta_{12} \sin^2\!\theta_{23}
 \frac{| m_1 e^{i\varphi_1} \!+ m_2 e^{i\varphi_2} |^2}{\Delta m_{21}^2} ,
\end{equation}
where $y_\tau$ is the $\tau$ Yukawa coupling, $\varphi_1$ and $\varphi_2$ are Majorana phases, and $\Delta m_{21}^2 \equiv m_2^2 - m_1^2$; all quantities are running parameters at the scale $\mu$.

The factor $(m_1 + m_2)^2/\Delta m_{21}^2$, to which the last factor on the right-hand side reduces for equal Majorana phases, makes the running large for nearly degenerate masses; for $m_1 = 0.05$~eV, it exceeds 100. This effect, which we call the degeneracy enhancement, reflects a general property of matrices: when two eigenvalues are nearly equal, a small correction to the matrix rotates the corresponding eigenvectors by a large angle (Sec.~\ref{sec:general-physical}). Claims of large running rely on this effect.

Observables do not depend on $\mu$, but the one-loop corrections from which they are computed do, through terms like $\ln(\mu^2/M^2)$, where $M$ is a mass or a momentum that flows through the loop. Equations~(\ref{equ:rge}) and (\ref{equ:theta12_running}) follow from this dependence: they describe how the parameters must change with $\mu$ to compensate it. Integrating them from $\mu_0$ to $\mu$ sums the leading logarithms. In practice, a calculation is truncated at a fixed order; the neglected terms contain higher powers of $\ln(\mu^2/M^2)$. Therefore, one chooses $\mu$ close to $M$, so that these terms are small. 

However, all the loops in an amplitude share the same $\mu$, so loops with $M$ far from $\mu$ keep large logarithms. For instance, the premise identifies $\mu$ with the momentum transfer, $Q$, which enters the vertex corrections. The logarithms of the neutrino self-energy are $\ln(\mu^2/M_X^2)$, where $M_X$ are the masses of the particles in its loop. Hence, with $\mu = Q$, the self-energy contains the logarithms $\ln(Q^2/M_X^2)$, which are large when $Q$ and $M_X$ differ widely. Section~\ref{sec:general-physical} shows how these logarithms enter the physical mass matrix; Sec.~\ref{sec:general-muQ} shows that they rotate $U(Q)$ back to the physical mixing.

\smallskip

\textit{In short:} the running mixing matrix, $U(\mu)$, is a parameter of the Lagrangian, not an observable. Its dependence on $\mu$ compensates that of the loop corrections, so $\mu$ is a bookkeeping device. The running is large only through the degeneracy enhancement.

\subsection{Physical mass matrix and physical mixing}
\label{sec:general-physical}

The neutrino masses that govern oscillations are the pole masses, \ie, the positions of the poles of the neutrino propagator, including its self-energy. The same propagator defines the physical mixing matrix, whose columns give the flavor content of the propagating states ($\hat\nu_1$, $\hat\nu_2$, $\hat\nu_3$ in vacuum).

At one loop, the neutrino self-energy arises from loops of particles with masses $M_X$ much larger than the neutrino masses, such as the $W$, $Z$, and Higgs bosons, or new states. For momenta $p^2 \ll M_X^2$, the neutrinos are described by an effective Lagrangian, obtained by integrating out these heavy particles~\cite{Appelquist:1974tg, Weinberg:1978kz}. It contains only the lighter fields; the loops of the heavy particles enter through the Wilson coefficients of its operators~\cite{Wilson:1969zs}. The terms of the effective Lagrangian that are quadratic in the neutrino field, which determine its propagator, are, in two-component notation~\cite{Chankowski:2001hx, Dreiner:2008tw},
\begin{eqnarray}
 \label{equ:two_point}
 && \nu^\dagger \bar\sigma \cdot p \left[ 1 + \Sigma_K(p^2) \right] \nu \nonumber \\
 && \quad - \, \tfrac{1}{2} \big\{ \nu^T \left[ m(\mu) + \Sigma_M(p^2) \right] \nu + {\rm h.c.} \big\} ,
\end{eqnarray}
where $\bar\sigma^\mu = (1, -\vec\sigma)$, with $\vec\sigma$ the Pauli matrices, and $\Sigma_K$ and $\Sigma_M$ are the one-loop corrections to the kinetic term and to the mass term, respectively, renormalized at scale $\mu$. In flavor space, $\Sigma_K$ is Hermitian and $\Sigma_M$ is symmetric. Since the neutrino masses are tiny compared to $M_X$, the self-energies can be evaluated at $p^2 = 0$. The kinetic correction, $\Sigma_K$, changes the normalization of the neutrino field. Rescaling the field as $\nu \to [1 - \frac{1}{2}\Sigma_K(0)] \nu$ restores the standard normalization and moves $\Sigma_K$ into the mass term. The result is the physical mass matrix~\cite{Chankowski:2001hx, Chankowski:2001mx}
\begin{eqnarray}
 \label{equ:meff}
 \meff
 &=&
 m(\mu) + \Sigma_M(0) \nonumber \\
 && - \, \frac{1}{2} \left[ \Sigma_K^T(0) \, m(\mu) + m(\mu) \, \Sigma_K(0) \right] ,
\end{eqnarray}
up to corrections of $\mathcal{O}(m_i^2/M_X^2)$ and of second order in the loop expansion. Its eigenvalues are the pole masses, $\hat m_i$, which, for $\mu$ near $M_X$, differ from the $m_i$ only by small loop corrections; elsewhere, we write the mass shell as $p^2 = m_i^2$ without distinguishing them. Its eigenvectors define the physical mixing matrix, $\Uhat$~\cite{Chankowski:2001mx}:
\begin{equation}
 \label{equ:uhat}
 \Uhat^T \meff \, \Uhat = {\rm diag}(\hat m_1, \hat m_2, \hat m_3) \;.
\end{equation}

\begin{figure*}[t!]
 \centering
 \includegraphics[width=\textwidth]{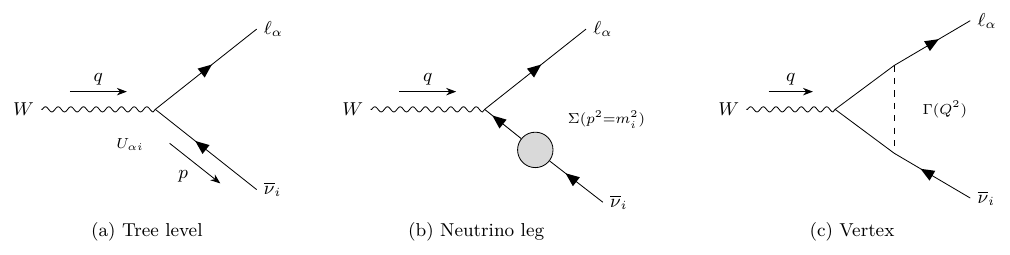}
 \caption{\textbf{\textit{One-loop structure of the charged-current amplitude.}} (a) Tree level. (b) Self-energy on the external neutrino leg, evaluated at the neutrino's own momentum, $p^2 = m_i^2$. This diagram contains the large rotation of the mixing for nearly degenerate masses, \ie, the physics behind the RG running of mixing. (c) Vertex correction. Apart from the self-energies of the $W$, which are the same for every flavor, the momentum transfer, $Q = |q^2|^{1/2}$, enters loop integrals only in diagrams of this type and in box diagrams.}
 \label{fig:cc_amplitude}
\end{figure*}

Since $\meff$ is physical, it is independent of $\mu$~\cite{Chankowski:2001hx, Chankowski:2001mx}. Its total derivative with respect to $\ln\mu$ vanishes, so to one loop
\begin{eqnarray}
 \label{equ:rge_from_selfenergy}
 \frac{dm}{d\ln\mu}
 &=&
 -\frac{\partial}{\partial \ln\mu}
 \bigg[ \Sigma_M(0) \nonumber \\
 && \qquad\quad - \, \frac{1}{2} \left( \Sigma_K^T(0) m + m \Sigma_K(0) \right) \bigg] .
\end{eqnarray}
This is the RG equation for $m(\mu)$, written in terms of the self-energy. In a schematic example, the self-energy contains only a kinetic correction, $\Sigma_K(0) = -c \ln(\mu^2/M_X^2)$, where $c$ is a small matrix in flavor space, \eg, proportional to $Y_e^\dagger Y_e$. By \equ{meff}, $\meff = m(\mu) + \frac{1}{2}(c^T m + m \, c) \ln(\mu^2/M_X^2)$. For $\meff$ to be independent of $\mu$, $m(\mu)$ must change as $dm/d\ln\mu \simeq -(c^T m + m \, c)$; this is its running. Since $c$ is not proportional to the identity, the running rotates the eigenvectors of $m(\mu)$. The second term of $\meff$ compensates this rotation, so $\Uhat$ does not change. That term vanishes at $\mu = M_X$. Hence, $\meff = m(M_X)$, whatever the value of $\mu$.

Two consequences follow. First, $\Uhat$ coincides, up to small finite corrections, with $U(\mu)$ evaluated at a scale $\mu$ near the masses of the particles in the self-energy, where the logarithms in $\Sigma_K$ and $\Sigma_M$ are small. When $\mu$ falls below the mass of a heavy particle, that particle is integrated out; its loops then stop contributing to the running. In the SM with massive neutrinos, this means that $\Uhat$ is $U(\mu \approx M_Z)$, up to small threshold corrections, \ie, the finite parts of the self-energy that remain after the heavy particles are integrated out.

Second, the degeneracy enhancement of \equ{theta12_running} reappears in $\Uhat$. As in first-order perturbation theory in quantum mechanics, a correction mixes two eigenvectors by an angle equal to the off-diagonal element of the correction divided by the difference between the two eigenvalues (for a Hamiltonian with eigenstates $|i\rangle$ and energies $E_i$, a correction $V$ turns $|j\rangle$ into $|j\rangle + \sum_{i \neq j} |i\rangle \langle i | V | j \rangle / (E_j - E_i)$). Thus, the eigenvectors of $\meff$ differ from those of $m(\mu)$ by a rotation with angles~\cite{Kniehl:1996bd, Chankowski:2001mx}
\begin{equation}
 \label{equ:enhancement}
 \epsilon_{ij}
 \simeq
 \frac{(\Sigma_M)_{ij} - \frac{1}{2}(m_i + m_j)(\Sigma_K)_{ij}}{m_j - m_i} \;,
 \quad i \neq j \;.
\end{equation}
Here, all matrices are written in the basis in which $m(\mu)$ is diagonal, with real, positive eigenvalues $m_i$. The first term in the numerator comes from the correction to the mass term; the second, from the correction to the kinetic term. Equation~(\ref{equ:enhancement}) holds for the real parts of the corrections. The imaginary parts enter with $1/(m_i + m_j)$, which is not enhanced. In the basis with real, positive $m_i$, the Majorana phases determine which parts of the corrections are real; this is the origin of the dependence on Majorana phases in \equ{theta12_running}. Equation~(\ref{equ:enhancement}) shows that the degeneracy enhancement, which makes the running large, enters $\Uhat$ through the neutrino self-energy, evaluated at the neutrino's own momentum, $p^2 = m_i^2 \simeq 0$.

Equations~(\ref{equ:meff}) and (\ref{equ:uhat}) are the standard definition of the mixing matrix in the on-shell renormalization scheme, in which masses and mixing are defined through the physical, on-shell particles~\cite{Denner:1990yz, Kniehl:1996bd, Kniehl:2006bs, Almasy:2009kn}; see Appendix~\ref{app:twopoint}.

\smallskip

\textit{In short:} the physical mixing is fixed by the neutrino propagator, including its self-energy, at $p^2 = m_i^2$; it does not depend on $\mu$. The self-energy there depends on the masses of the particles in its loops, but not on the momentum transfer or the neutrino energy.

\subsection{Production and detection amplitudes}
\label{sec:general-amplitude}

In an oscillation experiment, production, propagation, and detection form a single amplitude. The neutrino is exchanged between the production and detection vertices~\cite{Jacob:1961zz, Kobzarev:1980nk, Kobzarev:1981ra, Giunti:1993se}. At macroscopic distances, only neutrinos with $p^2$ very close to one of the $m_i^2$ contribute; contributions far from these poles of the propagator fall off rapidly with distance~\cite{Grimus:1996av, Beuthe:2001rc, Akhmedov:2010ms}. Hence, the propagating neutrino is on its mass shell, $p^2 = m_i^2$. The amplitude factorizes into the production amplitude of $\hat\nu_i$, its propagation, and its detection amplitude. The works in Table~\ref{tab:literature} use the same factorization, with on-shell mass eigenstates propagating between production and detection; see, \eg, \Refe~\cite{Babu:2021cxe}.

Figure~\ref{fig:cc_amplitude} shows the one-loop structure of the charged-current amplitude for the decay of a virtual $W$ boson, $W^{-\ast}(q) \to \ell_\alpha + \overline{\nu}_i(p)$. The same structure holds for the production and detection processes, which follow from this one by exchanging initial- and final-state particles. As in Sec.~\ref{sec:intro}, the momentum transfer is $Q = |q^2|^{1/2}$. There are four classes of corrections:
\begin{itemize}[leftmargin=*]
 \item \textit{Self-energies on the external neutrino leg} [\figu{cc_amplitude}(b)]. They depend on the neutrino four-momentum only through $p^2 = m_i^2$; in vacuum, Lorentz invariance forbids any dependence on the neutrino energy. The momentum transfer does not flow through them.

 \item \textit{Self-energies on the external charged-lepton leg.} They depend on the charged-lepton four-momentum only through $m_\ell^2$. They can rotate the charged-lepton states, and thereby the mixing, by angles of order an off-diagonal correction divided by a charged-lepton mass difference, as in \equ{enhancement}. Since the charged-lepton masses are very different from one another, these angles lack the degeneracy enhancement and are small.

 \item \textit{Self-energies of the $W$.} They depend on $q^2$, and hence on $Q$, but they are the same for every flavor.

 \item \textit{Vertex and box corrections} [\figu{cc_amplitude}(c)]. They depend on $Q$. Part of them also depends on flavor, because the particles in their loops can couple to each lepton flavor with a different strength, \eg, through the charged-lepton Yukawa couplings or through new interactions. Hence, these are the only corrections that depend both on flavor and on $Q$.
\end{itemize}
With these corrections, the amplitude for producing or detecting the state $\hat\nu_i$ with charged lepton $\ell_\alpha$ is~\cite{Denner:1990yz, Denner:1991kt, Kniehl:1996bd}
\begin{eqnarray}
 \label{equ:amplitude}
 \mathcal{A}_{\alpha i}(Q^2)
 &=&
 \frac{g}{\sqrt{2}}
 \sum_\beta
 \Big[ \left( 1 + F(Q^2) \right) \delta_{\alpha\beta} \nonumber \\
 && \qquad\quad + \, \Gamma_{\alpha\beta}(Q^2) + \Delta_{\alpha\beta} \Big]
 \Uhat_{\beta i} \;.
\end{eqnarray}
Here, $\hat\nu_i$ is a mass eigenstate, \ie, an eigenstate of the physical mass matrix, $\meff$. The function $F$ is the flavor-universal part of the corrections. The matrix $\Gamma$ is the flavor-dependent part of the vertex and box corrections. The external-leg corrections convert $U(\mu)$ into $\Uhat$. The matrix $\Delta$ is the flavor-dependent part of these corrections that remains after this conversion (Appendix~\ref{app:twopoint}). It is small and Hermitian, so it does not rotate the flavor states (a rotation needs an anti-Hermitian matrix). For example, the rescaling below \equ{two_point}, $\nu \to [1 - \frac{1}{2}\Sigma_K(0)]\nu$, acts on the neutrino field everywhere. In the mass term, it was absorbed into $\meff$, hence into $\Uhat$; in the charged-current vertex, it contributes $-\frac{1}{2}\Sigma_K(0)$ to $\Delta$.

In \equ{amplitude}, only $F$ and $\Gamma$ depend on $Q$: both $\Uhat$ and $\Delta$ come from the external-leg corrections, through which $Q$ does not flow. Therefore, the physical mixing, $\Uhat$, is the same at production and detection. Moreover, the large rotations due to the degeneracy enhancement arise only on the external neutrino leg (Sec.~\ref{sec:general-vertex}); they are contained in $\Uhat$, not in $\Gamma$.

\smallskip

\textit{In short:} production and detection see the same mixing, $\Uhat$, up to the flavor-dependent part of the vertex and box corrections, which is small (Sec.~\ref{sec:general-vertex}).

\subsection{Vertex corrections cannot reproduce running}
\label{sec:general-vertex}

The factor $1/(m_j - m_i)$ in \equ{enhancement} requires a neutrino near its mass shell. In diagrams, it arises from a self-energy on the external line of a neutrino $\nu_i$, followed by the propagator of another state, $\nu_j$, evaluated at $p^2 = m_i^2$: $1/(p^2 - m_j^2) = 1/(m_i^2 - m_j^2)$ (Appendix~\ref{app:twopoint}). This factor is large when $m_i \simeq m_j$; it turns a small correction into a large mixing of $\nu_i$ and $\nu_j$. In a vertex or box diagram, instead, an internal neutrino line carries a momentum far from the mass shell, with $|p^2|$ of order $Q^2$ or of order the squared masses in the loop. There, the propagators of the mass eigenstates are nearly equal, $1/(p^2 - m_i^2) \simeq 1/(p^2 - m_j^2) \simeq 1/p^2$, so no large factor appears. The neutrino masses enter only through tiny ratios, such as $m_i^2/M_X^2$. Hence, the flavor-dependent part of the vertex correction has the generic size~\cite{Bellazzini:2010gn}
\begin{equation}
 \label{equ:vertex_size}
 \Gamma_{\alpha\beta}(Q^2)
 \sim
 \frac{y_\alpha y_\beta^\ast}{16\pi^2} \, f\!\left( \frac{Q^2}{M_X^2} \right) .
\end{equation}
Here, $y_\alpha$ is the coupling of the new particles in the loop, of mass $M_X$, to $\ell_\alpha$ and $\nu_\alpha$. The function $f$ is of order unity; for $Q \gg M_X$, it grows at most as $\ln^2(Q^2/M_X^2)$~\cite{Denner:2000jv}.

Unlike the rotation angles in \equ{enhancement}, the vertex correction $\Gamma$ contains no factor $1/(m_j - m_i)$. In the first three examples of Sec.~\ref{sec:examples}, it is flavor-diagonal, tiny, or zero. In the MSSM (Sec.~\ref{sec:examples-mssm}), for flavor-universal superpartner masses, it is flavor-diagonal: it rescales the $\nu_\tau$ coupling but cannot rotate $\Uhat$. In the seesaw with a GeV-scale HNL (Sec.~\ref{sec:examples-hnl}), it is at most $4 \times 10^{-9}$ for the benchmarks of Table~\ref{tab:hnl_benchmarks}. In the seesaw with a light scalar (Sec.~\ref{sec:examples-model2}), it vanishes at one loop. Hence, vertex corrections cannot produce the rotation of the mixing the premise attributes to running.

\smallskip

\textit{In short:} corrections at the interaction point cannot mimic running. They lack the degeneracy enhancement, so they remain suppressed by a loop factor and couplings.

\subsection{What setting \texorpdfstring{$\mu = Q$}{mu = Q} does}
\label{sec:general-muQ}

We address in turn the three justifications of the choice $\mu = Q$ listed in Sec.~\ref{sec:literature}:
\begin{enumerate}[leftmargin=*, label=(\arabic*)]
 \item
  The first justification (``vertex corrections are evaluated with $\mu \approx Q$ to avoid large logarithms'') has two steps. The first step, evaluating the vertex correction at $\mu \approx Q$, is legitimate. The second step, concluding that the mixing in the process is $U(\mu = Q)$, is not. If one renormalizes at $\mu = Q$, the tree-level vertex carries $U(Q)$. The external neutrino leg then contributes $\Sigma_K(0)$ and $\Sigma_M(0)$ renormalized at $\mu = Q$. These contain the large logarithms $\ln(Q^2/M_X^2)$. Through the degeneracy enhancement, \equ{enhancement}, they rotate the columns of $U(Q)$ by large angles.

  The schematic example of Sec.~\ref{sec:general-physical} shows what this rotation does. There, $\meff = m(Q) + \frac{1}{2}(c^T m + m \, c) \ln(Q^2/M_X^2)$; the second term undoes the running of $m(\mu)$ from $M_X$ to $Q$, including the rotation of its eigenvectors. Thus, through \equ{meff}, the external leg rotates $U(Q)$ back to $\Uhat$. Hence, the running is not the independent source of corrections that the first justification assumes: the external leg cancels it. Keeping $U(Q)$ while dropping the logarithms on the external leg keeps the running without the term that undoes it, so the result depends on the arbitrary choice $\mu = Q$. The cleanest choice is $\mu \approx M_X$, for which the self-energy contains no large logarithms.

 \item
  The second justification (``the running of the mixing should come from the charged-current vertex'') also has two steps. The first step, treating production and detection separately because the neutrino propagates on shell, is legitimate (Sec.~\ref{sec:general-amplitude}). The second step, combining an on-shell propagator, renormalized at the neutrino masses, with a vertex that carries $U(Q)$, is not. The propagator and the vertex belong to the same amplitude, so they must be renormalized in the same scheme. If the propagator is renormalized on shell, the vertex carries $\Uhat$ (Sec.~\ref{sec:general-physical}). If instead $\mu = Q$ is used throughout, the self-energy on the external neutrino leg contains the logarithms that rotate $U(Q)$ back to $\Uhat$. Either way, the mixing is $\Uhat$; the vertex adds only the small $\Gamma(Q^2)$ (Sec.~\ref{sec:general-vertex}).

 \item
  The third justification (``a coupling is defined by an observable at momentum transfer $Q$''), which invokes the Gell-Mann--Low scheme, fails as well. Taken at face value, the scheme would define the mixing at $Q$ through the charged-current amplitude, \equ{amplitude}. In that amplitude, only the small $\Gamma(Q^2)$ depends on both flavor and $Q$ (Sec.~\ref{sec:general-amplitude}). Hence, a mixing defined in this way would depend on $Q$ only weakly. In practice, the works that invoke the scheme do not define the mixing through an amplitude. They set $\mu = Q$ in RG equations for the mixing parameters, which describe how Lagrangian parameters change with $\mu$~\cite{Antusch:2003kp, Antusch:2005gp}. That is a choice of $\mu$. As shown for the first justification, the external neutrino leg then contains the logarithms that rotate $U(Q)$ back to $\Uhat$.
\end{enumerate}

\smallskip

\textit{In short:} for any choice of $\mu$, including $\mu = Q$, the production and detection amplitudes contain the physical mixing, $\Uhat$, up to the small $\Gamma(Q^2)$. Using $U(Q)$ while dropping the logarithms on the external neutrino leg, which undo its running, is an inconsistent truncation.

\subsection{Consequences}
\label{sec:general-consequences}

In vacuum, the oscillation probability at a distance $L$ from the source is the standard one~\cite{Giunti:2007ry}, built from $\Uhat$ and the pole masses,
\begin{equation}
 \label{equ:probability}
 P_{\alpha\beta}(L, E)
 =
 \Big| \sum_i \Uhat_{\alpha i}^\ast \, \Uhat_{\beta i} \, e^{-i \hat m_i^2 L/2E} \Big|^2 ,
\end{equation}
up to the small, unenhanced corrections in \equ{amplitude}. With these corrections, the production and detection vertices carry $N_p \equiv (1 + \Gamma_p + \Delta)\Uhat$ and $N_d \equiv (1 + \Gamma_d + \Delta)\Uhat$ instead of $\Uhat$, where $\Gamma_p$ and $\Gamma_d$ are the values of $\Gamma$ at production and detection. The probability, normalized to the flux and the detection cross section, takes the form derived in \Refe~\cite{Antusch:2006vwa}, with $N_p$ and $N_d$ in place of a single matrix:
\begin{equation}
 \label{equ:probability_full}
 P_{\alpha\beta}(L, E)
 =
 \frac{\big| \sum_i (N_p)_{\alpha i}^\ast \, (N_d)_{\beta i} \, e^{-i \hat m_i^2 L/2E} \big|^2}{(N_p N_p^\dagger)_{\alpha\alpha} \, (N_d N_d^\dagger)_{\beta\beta}} .
\end{equation}
The flavor-universal $F$ [see \equ{amplitude}] cancels in it. For $\Gamma_p = \Gamma_d = \Delta = 0$, \equ{probability_full} reduces to \equ{probability}. The diagonal elements of $\Gamma_p$, $\Gamma_d$, and $\Delta$ also cancel~\cite{Falkowski:2019xoe}. The off-diagonal ones, which violate lepton flavor, change the probability by at most an amount of the order of these elements, except at $L = 0$, where the change is of second order in them. Hence, zero-distance effects~\cite{Langacker:1988up, Antusch:2006vwa} in the probability are even smaller.

In the SM, $\Gamma$ is flavor-diagonal at one loop, since the photon and the $Z$ and Higgs bosons couple diagonally to the charged leptons~\cite{Glashow:1976nt, Paschos:1976ay}. The off-diagonal elements of $\Delta$ come from the self-energy of the external charged lepton, a loop with a $W$ boson and a neutrino. Unitarity of the mixing cancels their part that does not depend on the neutrino masses, leaving tiny terms, of order $(g^2/16\pi^2)\,\Delta m_{31}^2/M_W^2 \approx 10^{-27}$, where $M_W$ is the $W$ mass. This is the Glashow--Iliopoulos--Maiani mechanism~\cite{Glashow:1970gm}; with a photon attached, the same loop gives $\mu \to e + \gamma$, which it makes unobservable in the SM~\cite{Petcov:1976ff, Bilenky:1977du, Marciano:1977wx, Lee:1977tib}. In matter, the standard Hamiltonian including the potential from coherent forward scattering on electrons, built from the same $\Uhat$ and pole masses, applies~\cite{Wolfenstein:1977ue, Mikheyev:1985zog}.

Beyond the SM, new particles that couple to two lepton flavors can generate larger off-diagonal elements, of the generic size of \equ{vertex_size}. The off-diagonal elements of $\Delta$ do not depend on $Q$; they have the same form as the effects of mixing with heavy neutrinos, which data bound to below about $10^{-3}$~\cite{Fernandez-Martinez:2016lgt}. Those of $\Gamma$ act as non-standard charged-current interactions at production and detection, which data bound to below about $10^{-2}$--$10^{-1}$~\cite{Biggio:2009nt, Farzan:2017xzy, Proceedings:2019qno}. Hence, even beyond the SM, the vertex corrections are small: for couplings of order unity, \equ{vertex_size} gives $|\Gamma_{\alpha\beta}| \sim 10^{-2}$, with no degeneracy enhancement to amplify it.

\begin{table*}[t!]
 \footnotesize
 \caption{\label{tab:comparison}\textbf{\textit{Two uses of the RG running of neutrino mixing.}} In the standard use, the running mixing matrix, $U(\mu)$, connects the parameters that a model fixes at a high scale to the physical mixing, $\Uhat$ (Sec.~\ref{sec:antusch}). The premise instead takes $U(\mu = Q)$ as the mixing seen at momentum transfer $Q$. Only the standard use is correct: every experiment, at any momentum transfer, measures the same $\Uhat$ (Sec.~\ref{sec:general-muQ}).}
 \begin{ruledtabular}
  \begin{tabular}{p{0.14\textwidth} p{0.40\textwidth} p{0.40\textwidth}}
   & Standard (\eg, \Refes~\cite{Antusch:2001ck, Antusch:2003kp, Antusch:2005gp}) & Premise of Table~\ref{tab:literature} \\
   \hline
   Question & High-scale model $\to$ low-energy observables & Does the mixing differ between processes at different $Q$? \\
   Meaning of $\mu$ & Bookkeeping scale & Momentum transfer of the process \\
   Output & One mixing matrix, $\Uhat$ & $U(Q_p) \neq U(Q_d)$ \\
   Status & Correct & Incorrect (Sec.~\ref{sec:general-muQ}) \\
  \end{tabular}
 \end{ruledtabular}
\end{table*}

Hence, none of the signatures proposed under the premise (Table~\ref{tab:literature}) arises:
\begin{enumerate}[leftmargin=*]
 \item
  \textit{Flavor change at zero or short distance from the source.---}At $L = 0$, $P_{\alpha\beta} = \delta_{\alpha\beta}$, up to terms of second order in the off-diagonal elements of $\Gamma_p$, $\Gamma_d$, and $\Delta$. At short distances, flavor change comes from oscillations, \eg, into sterile neutrinos, not from a difference between the mixing at production and at detection. Section~\ref{sec:data} compares the zero-distance flavor change that the premise predicts with short-baseline data.

 \item
  \textit{Shifts in the measured mixing parameters.---}Experiments at different neutrino energies, $E$, probe different momentum transfers. In deep-inelastic scattering on a nucleon of mass $m_N$, $Q^2 \leq 2 m_N E$; above $E \approx M_W^2/2m_N \approx 3$~TeV, the $W$ propagator keeps $Q$ near $M_W$. Thus, $Q$ ranges from a few MeV in reactor experiments, which detect neutrinos by inverse beta decay, through about 1~GeV in accelerator experiments and a few GeV for atmospheric neutrinos, to about $M_W$ for high-energy atmospheric and astrophysical neutrinos. All of these experiments measure the same $\Uhat$.

 \item
  \textit{Anomalous flavor composition of high-energy astrophysical neutrinos.---}At all energies, the flavor composition at Earth is $f_{\beta,\oplus} = \sum_\alpha f_{\alpha,{\rm S}} \sum_i |\Uhat_{\alpha i}|^2 |\Uhat_{\beta i}|^2$, where $f_{\alpha,{\rm S}}$ is the ratio of the flux of $\nu_\alpha + \bar{\nu}_\alpha$ to the total flux emitted at the neutrino sources. Section~\ref{sec:legit} discusses effects that can change it.
\end{enumerate}
In all three cases, the only dependence on the momentum transfer comes from the small $\Gamma(Q^2)$.

\smallskip

\textit{In short:} every experiment, at any baseline, energy, or momentum transfer, measures the same mixing, $\Uhat$. The signatures proposed under the premise do not arise: the RG running of the mixing is undetectable.

\section{Comparison with the standard treatment of RG running}
\label{sec:antusch}

\subsection{Same calculation, different question}
\label{sec:antusch-same}

Our result does not contradict any of the standard works on the RG running of neutrino masses and mixing~\cite{Babu:1993qv, Chankowski:1993tx, Casas:1999ac, Antusch:2001ck, Antusch:2001vn, Chankowski:2001mx, Antusch:2002rr, Antusch:2003kp, Antusch:2005gp, Ohlsson:2013xva}. Those works answer a different question: how are the parameters of a model at a high scale related to the masses and mixing measured at low energies? In the words of \Refe~\cite{Antusch:2001ck}, it is ``essential to evolve the predictions to low energies'' before comparing them with experiment. Those that solve the RG equations run $U(\mu)$ from the high scale down to $\mu \sim M_Z$ and identify $U(M_Z)$ with the mixing measured in oscillation experiments. As shown in Sec.~\ref{sec:general-physical}, this is correct: the physical mixing, $\Uhat$, is $U(\mu)$ evaluated at the masses of the particles in the neutrino self-energy, up to small threshold corrections; below these masses, $U(\mu)$ no longer runs.

Below their masses, heavy particles are integrated out~\cite{Appelquist:1974tg}. Their effects are absorbed into the coefficients of the effective Lagrangian (Sec.~\ref{sec:general-physical}), a procedure called matching~\cite{Antusch:2002rr, Antusch:2005gp}; the scale at which it is done is the matching scale. Beyond tree level, matching produces threshold corrections, \ie, finite shifts of the parameters~\cite{Chun:1999vb, Chankowski:2001hx, Chankowski:2001mx}; these are the finite parts of $\Sigma_K$ and $\Sigma_M$ in \equ{meff}. As \Refe~\cite{Chankowski:2001mx} notes, including them makes the predicted observables independent of the matching scale.

Table~\ref{tab:comparison} shows the two uses of the RG running of neutrino mixing: the standard one and the premise. None of the standard works claims that an experiment at energy $E$, or momentum transfer $Q$, sees $U(\mu = E)$ or $U(\mu = Q)$. That claim is what the premise adds.

\subsection{What RG running is good for}
\label{sec:antusch-use}

What runs is a parameter of the Lagrangian, not the physical mixing: the running mixing matrix, $U(\mu)$, with its running angles, $\theta_{ij}(\mu)$ (Sec.~\ref{sec:general-running}). At a generic scale, these are intermediate quantities of the calculation. Only for $\mu \lesssim M_X$ do the running angles coincide, up to small threshold corrections (Sec.~\ref{sec:antusch-same}), with the physical angles, which every experiment measures.

The reason is \equ{meff}. The physical angles parametrize the eigenvectors of $\meff$; the running angles parametrize those of $m(\mu)$ alone. The corrections that separate the two contain logarithms of $\mu/M_X$. At $\mu \approx M_X$, these logarithms vanish, so the two sets of eigenvectors agree up to a small, finite threshold correction. Above $M_X$, $m(\mu)$ has run away from $\meff$, so its eigenvectors differ from the physical ones; the large logarithms in the self-energy compensate the difference. Below $M_X$, the heavy particles are integrated out, so $U(\mu)$ stops running and stays equal to $\Uhat$, up to the threshold correction (Sec.~\ref{sec:general-physical}).

\begin{figure}[t!]
 \centering
 \includegraphics[width=\columnwidth]{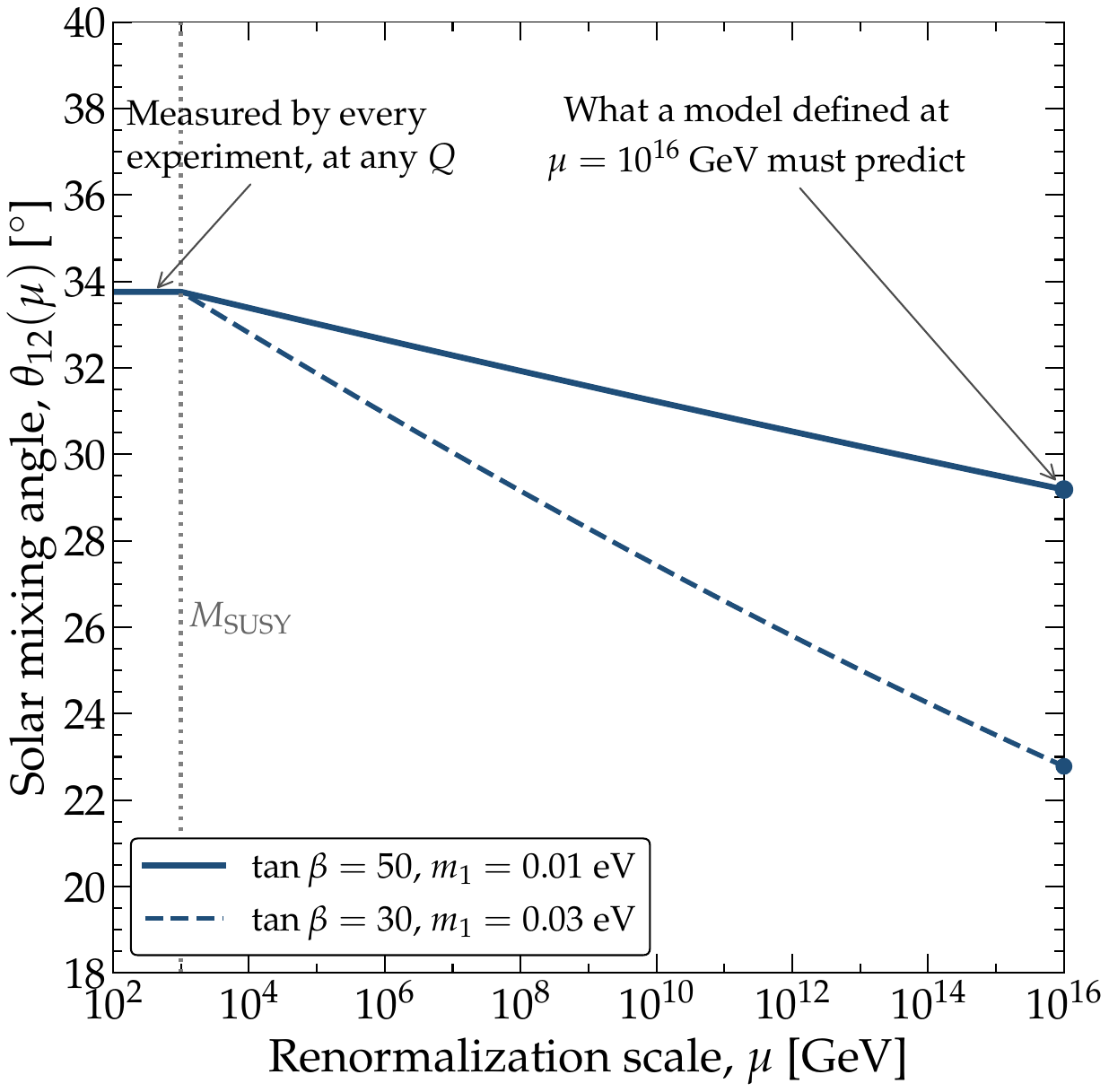}
 \caption{\textbf{\textit{What RG running is for.}} Running of the solar mixing angle, $\theta_{12}(\mu)$, a parameter of the Lagrangian, in the MSSM ($M_{\rm SUSY} = 1$~TeV, vanishing Majorana phases), for two choices of $\tan\beta$ and $m_1$. The running maps the input that a model must set at a high scale (right) onto the physical angle that every experiment measures (left). The horizontal axis is the renormalization scale, not the energy or the momentum transfer of any experiment.}
 \label{fig:rg_use}
\end{figure}

The running is needed when a model fixes the parameters of its Lagrangian at a scale far above $M_X$. This scale, $\mu_0$ in \equ{rge}, is the one at which the model generates the flavor structure of the neutrino mass matrix, \eg, the scale of a flavor symmetry or of the seesaw.

Figure~\ref{fig:rg_use} shows an example, in which a flavor symmetry at $\mu \simeq 10^{16}$~GeV fixes $\theta_{12}(10^{16}~{\rm GeV})$. Between $M_{\rm SUSY} = 1$~TeV, the mass scale of the superpartners, and $10^{16}$~GeV, the theory is the MSSM with $\tan\beta = 50$, where $\tan\beta$ is the ratio of the vacuum expectation values of its two Higgs doublets. This value makes the running large, but values this large are now disfavored: searches at the Large Hadron Collider for heavy Higgs bosons decaying into $\tau$ pairs exclude $\tan\beta > 21$ for heavy neutral Higgs bosons of mass 1.5~TeV in a benchmark scenario of the MSSM~\cite{ATLAS:2020zms}. We solve \equ{rge} at one loop, with $C = 1$ and $Y_e \simeq {\rm diag}(0, 0, y_\tau)$, holding $y_\tau$ fixed, as in Sec.~\ref{sec:examples-mssm}~\cite{Antusch:2003kp}. Since the model fixes its parameters at $10^{16}$~GeV, the neutrino self-energy, renormalized at that scale, contains large logarithms of $10^{16}~{\rm GeV}/M_{\rm SUSY}$; the running sums them. They shift the physical angle away from the input of the model. For $m_1 = 0.01$~eV, the model must set $\theta_{12}(10^{16}~{\rm GeV}) = 29.2^\circ$ to produce the measured physical angle, $33.8^\circ$. For $\tan\beta = 30$ and $m_1 = 0.03$~eV, the required input would instead be $22.8^\circ$. This is how the running enters tests of flavor symmetries, of grand unification, and of leptogenesis~\cite{Antusch:2003kp, Antusch:2005gp, Ohlsson:2013xva}.

The running in this example plays the same role as the one-loop correction, $\delta m$, to the light-neutrino mass matrix from the heavy neutral lepton in the HNL example of Sec.~\ref{sec:examples-hnl}: it turns the input of a model into the physical mixing. In both cases, the correction comes from the neutrino self-energy. Its size depends on mass parameters of the Lagrangian, \ie, on $10^{16}$~GeV and $M_{\rm SUSY}$ here, or on $M$ and on the masses of the $Z$ and Higgs bosons, $M_Z$ and $M_h$, for the HNL. It does not depend on the momentum transfer of any process. Every experiment, from reactors to IceCube, measures the physical angle, \ie, the left end of the curves in \figu{rg_use}.

The premise of Table~\ref{tab:literature} instead treats $\theta_{ij}(\mu)$ at the momentum transfers of production and detection, $\mu = Q_p$ and $\mu = Q_d$, as two physical mixings. However, the momentum transfer of a process is not a scale of the theory, \ie, not a mass parameter of the Lagrangian, such as the mass of a particle or the scale at which a model fixes its parameters. It does not appear in the neutrino self-energy, which fixes the mixing.

A model can fix its parameters at the momentum transfer of neutrino production if it generates the flavor structure through new particles with masses $M_X \sim m_\pi$, the momentum transfer of pion decay. The premise then appears to hold at production: $U(m_\pi)$, the input of the model, equals the physical mixing, up to small threshold corrections, because $U(\mu)$ stops running below $M_X$. It still fails at detection: at $Q_d \gg m_\pi$, $U(Q_d)$ has run away from the input, but the physical mixing has not, because the self-energy compensates the running (Sec.~\ref{sec:general-muQ}). Hence, every experiment measures the physical mixing; that the scale of the model coincides with the momentum transfer of pion decay plays no role.

\subsection{A parallel in the quark sector}
\label{sec:antusch-quarks}

The quark mixing matrix, the Cabibbo--Kobayashi--Maskawa (CKM) matrix, runs, as a parameter of the Lagrangian, above the electroweak scale~\cite{Babu:1987im}. As \Refe~\cite{Babu:2021cxe} notes, this running has never been observed. That work notes that, below the electroweak scale, the CKM matrix does not run unless light new particles couple to quarks. Yet the running would not be observable as a dependence on the momentum transfer even with such particles.

As for neutrinos, the CKM matrix that enters amplitudes with on-shell quarks is fixed by the quark propagators, including their self-energies, on the external lines. This is the on-shell definition of the CKM matrix used in calculations of electroweak radiative corrections~\cite{Denner:1990yz, Gambino:1998ec, Kniehl:2006bs}. The CKM matrix defined in this way does not depend on the momentum transfer of the process; apart from flavor-universal corrections, only vertex and box corrections do. In practice, CKM elements are extracted from processes with momentum transfers that range from a few MeV, in nuclear beta decays, to $M_W$, in decays of $W$ bosons and top quarks. Global fits combine all of them as elements of a single matrix~\cite{ParticleDataGroup:2024cfk}. 

Hence, as for neutrinos, the running of the CKM matrix is not observable as a dependence on the momentum transfer, whatever its size. The quark and neutrino sectors differ only in how large the running of the Lagrangian parameter can be: quark masses are hierarchical, so the CKM matrix runs without the degeneracy enhancement that can make the running of neutrino mixing large.

\section{Worked examples}
\label{sec:examples}

\begin{figure*}[t!]
 \centering
 \includegraphics[width=\textwidth]{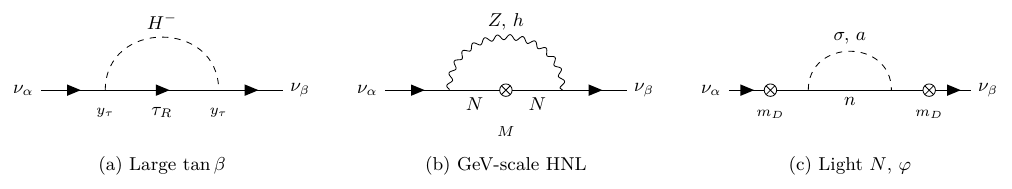}
 \caption{\textbf{\textit{Neutrino self-energies responsible for the running of mixing in our examples.}} (a) MSSM at large $\tan\beta$: loop of a charged Higgs boson and a right-chiral $\tau$ lepton, $\tau_R$, plus superpartner loops (Sec.~\ref{sec:examples-mssm}). (b) Seesaw with a GeV-scale Majorana HNL: $Z$ and Higgs loops with the HNL, $N$, which flips chirality through its mass, $M$ (Sec.~\ref{sec:examples-hnl}). (c) Seesaw with a light lepton-number-breaking scalar, $\varphi = (v_\varphi + \sigma + ia)/\sqrt{2}$: the scalar couples only to the sterile states, $n$; the active neutrino turns into a sterile state through the Dirac mass, $m_D$ (Sec.~\ref{sec:examples-model2}). Crossed circles mark mass insertions. In all cases, the loop is on the neutrino line, so it depends on $p^2 = m_i^2$, not on the momentum transfer.}
 \label{fig:selfenergies}
\end{figure*}

We now work out four explicit cases, chosen to match the settings of the works in Table~\ref{tab:literature}. In each one, we identify the origin of the running of the mixing, show that it comes from the neutrino self-energy, and estimate the corrections at the interaction point, the only flavor-dependent corrections that can depend on $Q$. Appendix~\ref{app:numerics} lists the numerical inputs.

Figure~\ref{fig:selfenergies} shows the neutrino self-energies responsible for the running of the mixing in three of these cases. In all of them, the loop is on the neutrino line, so it depends on $p^2 = m_i^2$, not on $Q$.

\subsection{MSSM at large \texorpdfstring{$\tan\beta$}{tan beta}}
\label{sec:examples-mssm}

This is the setting of \Refe~\cite{Bustamante:2010bf}. Above $M_{\rm SUSY}$, the mass scale of the superpartners, the Weinberg operator runs with $C = 1$ in \equ{rge}. The $\tau$ Yukawa coupling is enhanced: $y_\tau \simeq \sqrt{2} m_\tau / (v \cos\beta) \simeq 0.5$ for $\tan\beta = 50$, the value used in \Refe~\cite{Bustamante:2010bf}, now disfavored (Sec.~\ref{sec:antusch-use}). Here, $m_\tau$ is the $\tau$ mass and $v = 246$~GeV is the electroweak vacuum expectation value. With $Y_e \simeq {\rm diag}(0, 0, y_\tau)$, $y_\tau$ held fixed, and $\mu_0 = M_{\rm SUSY}$, the solution of \equ{rge} is $\kappa(t) = D \, \kappa(0) \, D$, up to a flavor-universal factor, with $D = {\rm diag}(1, 1, e^{y_\tau^2 t/16\pi^2})$.

\begin{figure*}[t!]
 \centering
 \includegraphics[width=\textwidth]{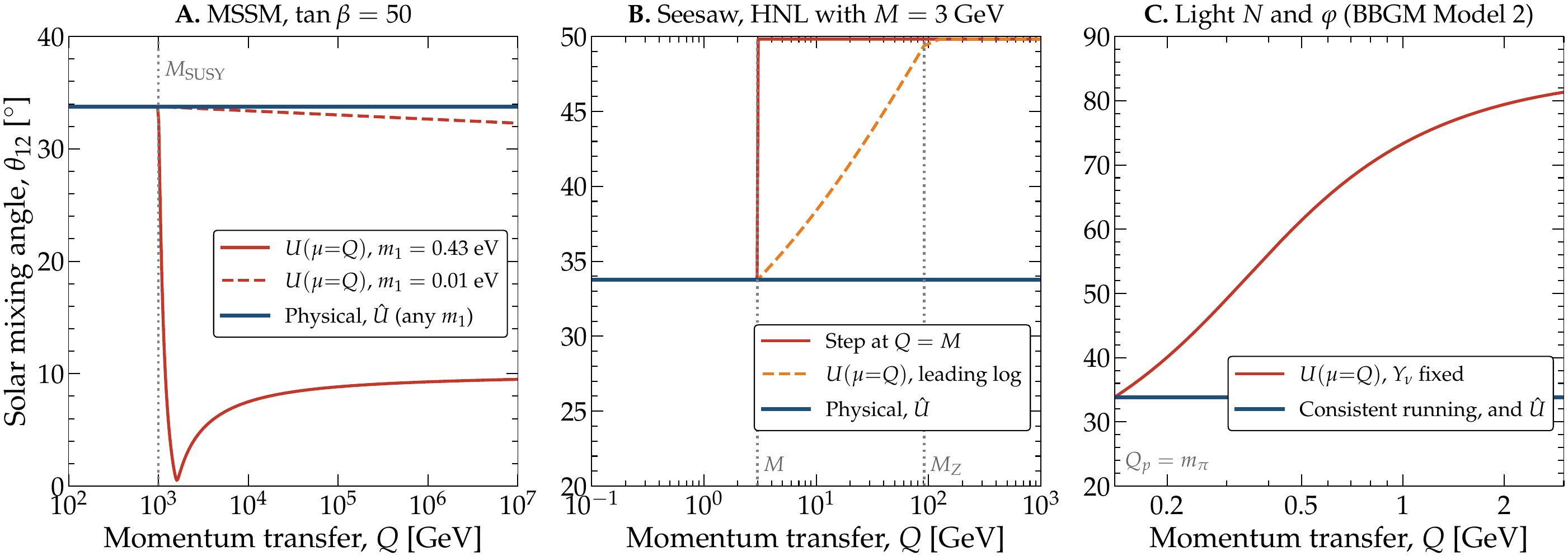}
 \caption{\textbf{\textit{Mixing at detection as a function of the momentum transfer: premise \textit{vs.}~physical.}} We show the solar mixing angle, $\theta_{12}$, that a detection process at momentum transfer $Q$ would see under the premise, and the physical angle. \textit{A.} MSSM with $\tan\beta = 50$ and $M_{\rm SUSY} = 1$~TeV: running of the Weinberg operator [\equ{rge}], with vanishing Majorana phases, for $m_1 = 0.43$~eV (as in \Refe~\cite{Bustamante:2010bf}) and $0.01$~eV. \textit{B.} Seesaw with an HNL of mass $M = 3$~GeV coupled to $\nu_\mu$ (benchmark B of Table~\ref{tab:hnl_benchmarks}): the step at $Q = M$ assumed in \Refe~\cite{Barenboim:2026mvn} and $U(\mu = Q)$ at leading-logarithmic accuracy. \textit{C.} Model~2 of \Refe~\cite{Babu:2021cxe} (BBGM), with the parameters of its benchmark BP2 (defined there for Model~1): running with the Dirac Yukawa matrix, $Y_\nu$, held fixed, as in that work, and with all couplings run consistently. In every case, the physical mixing does not depend on $Q$.}
 \label{fig:running_examples}
\end{figure*}

Figure~\ref{fig:running_examples} shows, for three of our cases, the solar mixing angle, $\theta_{12}$, that detection at momentum transfer $Q$ would see under the premise, together with the physical angle. Panel~A shows the resulting $\theta_{12}(\mu = Q)$ in the MSSM, with $M_{\rm SUSY} = 1$~TeV. For $m_1 = 0.43$~eV---the value used in \Refe~\cite{Bustamante:2010bf}, now excluded by cosmology~\cite{Planck:2018vyg, DESI:2024mwx}---the angle drops from $33.8^\circ$ to $0.5^\circ$ at $Q = 1.6$~TeV and is $7.5^\circ$ at 10~TeV. For $m_1 = 0.01$~eV, it changes by $0.4^\circ$ over the same range. Under the premise, the larger change would shift the flavor composition at Earth, $(f_e, f_\mu, f_\tau)_\oplus$, of neutrinos from pion decay detected at $Q_d = 10$~TeV from $(0.326, 0.343, 0.330)_\oplus$ to $(0.302, 0.376, 0.323)_\oplus$.

None of this is physical. In the MSSM, the Weinberg operator is a term of the superpotential, the function from which the Yukawa interactions of a supersymmetric theory follow. Supersymmetry forbids loop corrections to the operator itself, a result known as the non-renormalization theorem~\cite{Grisaru:1979wc}. Therefore, its running, \equ{rge}, comes entirely from the self-energies of the lepton and Higgs doublets, which change the normalization of their fields (wave-function renormalization). The flavor-dependent part of the running, $\propto Y_e^\dagger Y_e$, comes from the self-energy of the lepton doublet with a loop of a charged Higgs boson and a right-chiral $\tau$ lepton, $\tau_R$, and from the corresponding loop of their superpartners [\figu{selfenergies}(a)]. On an external neutrino, this self-energy is evaluated at $p^2 \simeq 0$, with loop masses of order $M_{\rm SUSY}$. It is the $\Sigma_K$ of \equ{two_point} for this model. In $\meff$, its logarithms of $\mu/M_{\rm SUSY}$ compensate the running of $m(\mu)$, as in Sec.~\ref{sec:general-physical}. Hence, the physical mixing, $\Uhat$, is $U(\mu \approx M_{\rm SUSY})$, up to finite supersymmetric threshold corrections~\cite{Chun:1999vb, Chankowski:2001hx}.

The only corrections that depend both on flavor and on $Q$ are the vertex and box corrections (Sec.~\ref{sec:general-amplitude}). In the MSSM, they come from loops with the charged and neutral Higgs bosons and with superpartners, such as higgsinos and sleptons. For flavor-universal superpartner masses, their flavor dependence enters only through the charged-lepton Yukawa couplings, \ie, the $y_\alpha$ of \equ{vertex_size}, of which only $y_\tau$ is sizable. Hence, they affect only the $\tau$ flavor: for $Q > M_{\rm SUSY}$, they rescale the $\nu_\tau$ coupling by an amount of order $(y_\tau^2/16\pi^2)\ln(Q^2/M_{\rm SUSY}^2)$, \ie, about $0.7\%$ at $Q = 10$~TeV and $2\%$ at $Q = 1$~PeV. They do not rotate the mixing. Moreover, deep-inelastic scattering of the high-energy astrophysical neutrinos that are the subject of \Refe~\cite{Bustamante:2010bf} is dominated by $Q \sim M_W$, below $M_{\rm SUSY}$, where the $Q$-dependence of these corrections is suppressed by $Q^2/M_{\rm SUSY}^2$. Reference~\cite{Bustamante:2010bf} found that averaging over the momentum transfer erases the effect of the running, unless only events with $Q^2 \gtrsim 10^7$~GeV$^2$ are kept; together with the low flux of ultra-high-energy neutrinos, this yields a negligible event rate. That conclusion stands, for a stronger reason: the effect is absent.

\smallskip

\textit{In short:} in the MSSM, the running of $U(\mu)$ can be large, but it is not observable as a dependence on the momentum transfer of the process. The only effect that depends both on flavor and on $Q$ is a percent-level change of the $\nu_\tau$ coupling.

\subsection{Seesaw with a GeV-scale heavy neutral lepton}
\label{sec:examples-hnl}

This is the setting of \Refe~\cite{Barenboim:2026mvn}. A Majorana HNL, $N$, with mass $M$ and mixing $U_{\alpha N}$ with the flavor states $\nu_\alpha$ contributes to the light-neutrino mass matrix at tree level, $m^{(N)}_{\alpha\beta} = -M U_{\alpha N} U_{\beta N}$. At one loop, it contributes through $Z$ and Higgs exchange [\figu{selfenergies}(b)]~\cite{Pilaftsis:1991ug, Grimus:2002nk, Lopez-Pavon:2015cga, Moffat:2018smo, Bolton:2019pcu}:
\begin{eqnarray}
 \label{equ:delta_m_hnl}
 \delta m_{\alpha\beta}
 &=&
 \frac{M^3 U_{\alpha N} U_{\beta N}}{16\pi^2 v^2}
 \left[ \frac{3 M_Z^2}{M^2 - M_Z^2} \ln\frac{M^2}{M_Z^2} \right. \nonumber \\
 && \left. \qquad\qquad\qquad + \frac{M_h^2}{M^2 - M_h^2} \ln\frac{M^2}{M_h^2} \right] \;.
\end{eqnarray}
For $M \ll M_Z$, the bracket is $3\ln(M_Z^2/M^2) + \ln(M_h^2/M^2) > 0$, so the loop has the opposite sign to the tree-level term. Their ratio, $R(M) \equiv \delta m / m^{(N)}$, is $-3.8 \times 10^{-6}$ for $M = 1$~GeV, $-2.6 \times 10^{-5}$ for 3~GeV, and $-1.9 \times 10^{-4}$ for 10~GeV.

For couplings within reach of the planned Search for Hidden Particles (SHiP) experiment~\cite{SHiP:2015vad, Alekhin:2015byh}, the tree-level contribution of $N$ is large, $M |U_{\alpha N}|^2 \sim 10^2$--$10^3$~eV, far above the light-neutrino masses. Therefore, another state, a partner of $N$ with mass $M^\prime$, must cancel it by contributing $-m^{(N)}$ at tree level. However, their loop contributions cancel only if the two masses are equal~\cite{Lopez-Pavon:2015cga, Jones-Perez:2023med}: the loop of $N$ contributes $R(M)\, m^{(N)}$, while that of the partner contributes $-R(M^\prime)\, m^{(N)}$. Since $|R(M)|$ grows with $M$, the loop of the partner is negligible if the partner is much lighter than $N$, and dominant if it is heavier. We assume a much lighter partner, so that the one-loop correction to the light-neutrino mass matrix is the loop of $N$, $\delta m$ in \equ{delta_m_hnl}.

Table~\ref{tab:hnl_benchmarks} shows the benchmarks of \Refe~\cite{Barenboim:2026mvn}, all within reach of SHiP. For them, $\delta m$ is a few meV, comparable to $m_2 - m_1$~\cite{Lopez-Pavon:2015cga}. The physical mass matrix, $\meff$ (Sec.~\ref{sec:general-physical}), is $m_{\rm tree} + \delta m$, where $m_{\rm tree}$ is the tree-level mass matrix, including the contributions of $N$ and of its partner. Its eigenvectors, $\Uhat$, differ from those of $m_{\rm tree}$ alone, $U_{\rm tree}$, by up to $0.27$ in $|U_{\alpha i}|^2$.

Reference~\cite{Barenboim:2026mvn} assumed that detection at $Q > M$ with high-energy astrophysical neutrinos can ``resolve'' the HNL, so that the mixing seen at detection differs from the mixing at $Q < M$ by the effect of $\delta m$, with a step at $Q = M$. It treated $\delta m$ as a threshold correction, distinct from RG running: it regarded the kinks that RG running shows in the $\overline{\rm MS}$ scheme as artifacts of a mass-independent scheme, in which heavy particles are integrated out abruptly at their mass.

The threshold correction is indeed physical: it is a one-time contribution to $\meff$. However, it is not separate from RG running. In the language of Sec.~\ref{sec:antusch-same}, $\delta m$ is the threshold correction that matching produces when $N$ is integrated out at $\mu \simeq M$; it contains the logarithms that the running from $M_Z$ to $M$ sums (Appendix~\ref{app:hnl}). Treating the matching scale, $\mu \simeq M$, as the momentum transfer of the process is the same step as setting $\mu = Q$. By the criterion of \Refe~\cite{Barenboim:2026mvn} itself, the resulting step at $Q = M$ is an artifact of the kind that it attributed to mass-independent schemes.

With RG running, as in the other works in Table~\ref{tab:literature}, the mixing at detection, $U(\mu = Q)$, would change gradually with $Q$, not in a step. At leading-logarithmic accuracy, the running builds up $\delta m$ as $\mu$ decreases from $M_Z$ and $M_h$ to $M$. Hence, at $\mu = Q$, only the part of $\delta m$ built up above $Q$ is present. This part shrinks as $Q$ grows: for $M = 3$~GeV, it is $46\%$ of $\delta m$ at $Q = 20$~GeV. Figure~\ref{fig:running_examples}B shows this gradual transition together with the step.

Neither the step nor the gradual transition is physical. The loop that generates $\delta m$ [\figu{selfenergies}(b)] is a self-energy of the neutrino: it is on the neutrino line and is evaluated at the neutrino's own momentum, $p^2 \simeq 0$. Hence, $\delta m$ is part of the physical mass matrix, $\meff$ [\equ{meff}], whose eigenvectors define the physical mixing, $\Uhat$. The momentum transfer, $Q$, is carried by the $W$ and does not flow through this loop. Therefore, detection at every $Q$, above or below $M$, sees the same $\Uhat$, which includes the effect of $\delta m$. The only effects of $N$ dependent on flavor and $Q$ are vertex and box corrections, $\sim (g^2/16\pi^2)|U_{\alpha N}|^2 \lesssim 4 \times 10^{-9}$ for the Table~\ref{tab:hnl_benchmarks} benchmarks.

Whether the HNL acts as a separate state, \ie, whether it is resolved, depends on the four-momentum of the neutrino, not on $Q$. A neutrino exchanged between two interaction points within the same process is virtual: its four-momentum squared, $p^2$, need not equal a mass squared. If $|p^2| \gtrsim M^2$, the HNL contributes as a separate state. In neutrinoless double beta decay, for example, the neutrino exchanged between two nucleons has $|p^2| \sim (100~{\rm MeV})^2$, so an HNL with $M \lesssim 100$~MeV is resolved~\cite{Bolton:2019pcu}. In contrast, the neutrinos that oscillation experiments and neutrino telescopes detect travel macroscopic distances, so they are on their mass shell, $p^2 = m_i^2 \ll M^2$, at every $Q$ (Sec.~\ref{sec:general-amplitude}). For them, the HNL enters through $\delta m$, not as a separate state.

\smallskip

\textit{In short:} the HNL shifts the physical mixing through $\delta m$, which does not depend on $Q$. Detection at $Q > M$ does not resolve the HNL: every experiment, at any $Q$, sees the shifted mixing, with no step at $Q = M$.

\subsection{Seesaw with a light scalar}
\label{sec:examples-model2}

This is Model~2 of \Refe~\cite{Babu:2021cxe}: a type-I seesaw, in which sterile neutrinos generate the light-neutrino masses, with light sterile neutrinos and a light scalar singlet, $\varphi$:
\begin{equation}
 \label{equ:model2_lagrangian}
 -\mathcal{L} = \overline{L} Y_\nu \tilde H N_R + \varphi \, \overline{N_R^c} Y_N N_R + {\rm h.c.} \;,
\end{equation}
where $L$ is the lepton doublet, $\tilde H = i\sigma_2 H^\ast$ is the conjugate of the Higgs doublet, $N_R$ are the sterile neutrinos, and $Y_\nu$ and $Y_N$ are Yukawa matrices. The scalar has the vacuum expectation value $\langle \varphi \rangle = v_\varphi/\sqrt{2}$, which gives the sterile neutrinos the masses $M = \sqrt{2} v_\varphi Y_N$ in our normalization. The light-neutrino mass matrix follows from the seesaw formula, $M_\nu \propto m_D M^{-1} m_D^T \propto Y_\nu Y_N^{-1} Y_\nu^T$, with Dirac masses $m_D = Y_\nu v/\sqrt{2}$.

Reference~\cite{Babu:2021cxe} runs $Y_N$ with
\begin{equation}
 \label{equ:beta_yn}
 16\pi^2 \frac{dY_N}{d\ln\mu} = 4 Y_N \left[ Y_N^2 + \tfrac{1}{2}{\rm Tr}(Y_N^2) \right] ,
\end{equation}
with $\mu = Q$. It keeps $Y_\nu$ fixed and diagonalizes $M_\nu(\mu = Q)$. Reference~\cite{Babu:2021cxe} shows numerical results only for its Model~1, a variant in which the light-neutrino masses arise at one loop, \ie, a scotogenic model, discussed below. We use Model~1 to validate our implementation: with its benchmark BP2, we recover the ratio $\theta_{12}(2.5~{\rm GeV})/\theta_{12}(Q_p) = 0.490$, read off its Fig.~8. In Model~2, which we analyze here, the same parameter values, with $Y_\nu$ fixed, give the $U(\mu = Q)$ curve in \figu{running_examples}C: $\theta_{12}$ grows from $33.8^\circ$ at $Q = m_\pi$ to $80.6^\circ$ at $2.5$~GeV, instead of decreasing as in Model~1.

\smallskip

\textit{The amplitude.---}We write the scalar as $\varphi = (v_\varphi + \sigma + ia)/\sqrt{2}$, where $\sigma$ and $a$ are real fields, with $a$ the Goldstone boson of the spontaneously broken lepton number, the majoron~\cite{Chikashige:1980ui, Schechter:1981cv}. At tree level, apart from a small mixing of $\sigma$ with the Higgs boson, $\sigma$ and $a$ couple only to the sterile states. They do not couple to the charged leptons, to the $W$, or to quarks. Therefore, any one-loop diagram involving $Y_N$ has the scalar attached at both ends to the neutrino line. It is a self-energy of the external neutrino [\figu{selfenergies}(c)]. No vertex diagram exists at this order. The effect of the self-energy is to replace each sterile mass, $M_k$, by
\begin{equation}
 \label{equ:meff_model2}
 M_k^{\rm eff}
 =
 M_k \left[ 1 + \frac{2 y_k^2}{16\pi^2} \, \frac{m_\sigma^2}{M_k^2 - m_\sigma^2} \ln\frac{M_k^2}{m_\sigma^2} \right]
\end{equation}
in the seesaw formula, where $y_k$ are the eigenvalues of $Y_N$, $M_k$ are the corresponding sterile masses, $m_\sigma$ is the mass of $\sigma$, and $a$ is taken to be massless (Appendix~\ref{app:model2}). The correction is finite and depends on the masses inside the loop only. The physical mixing is $Q$-independent.

\smallskip

\textit{The running, done consistently.---}We reproduce the $\beta$ function of \Refe~\cite{Babu:2021cxe}, \equ{beta_yn}, exactly. The one-loop correction to the $\varphi N N$ vertex has no ultraviolet divergence: lepton-number conservation forbids it for massless sterile neutrinos, so it must contain an insertion of the sterile mass, which makes it finite. Therefore, \equ{beta_yn} comes only from the self-energies of the fields, \ie, from the running of their normalization (wave-function renormalization). The rates of this running, called anomalous dimensions, are $\gamma_{n_k} = 2 y_k^2/16\pi^2$ for the sterile states, $n_k$, and $\gamma_\varphi = 2{\rm Tr}(Y_N^2)/16\pi^2$ for the scalar (Appendix~\ref{app:model2}).

The seesaw formula is a sum of terms, one for each sterile state, $n_k$, proportional to $(m_D)_{\alpha k} (m_D)_{\beta k}/M_k$. Each Dirac mass is the coefficient of a term in the Lagrangian with one sterile field. Each sterile mass is the coefficient of a term with two. Hence, the running of the normalization of $n_k$ makes $(m_D)_{\alpha k}$ run at the rate $\gamma_{n_k}$ and $M_k$ at the rate $2\gamma_{n_k}$. In each term of the seesaw formula, these rates cancel: $\gamma_{n_k} + \gamma_{n_k} - 2\gamma_{n_k} = 0$. Below the electroweak scale, the rest of the running rescales all terms equally, so the mixing does not run (Appendix~\ref{app:model2}). Thus, the divergent parts of the self-energies, which drive the running, cancel in the mixing; only the finite correction of \equ{meff_model2} remains.

The difference from \Refe~\cite{Babu:2021cxe} lies in which quantities are allowed to run. That work includes the running of $M_k$, through $Y_N$, but not that of $m_D$: it holds $Y_\nu$ fixed, on the grounds that the Higgs boson is integrated out below the electroweak scale. However, $m_D = Y_\nu v/\sqrt{2}$ is the coefficient of a term with the sterile field, which is light and is not integrated out. Therefore, $m_D$ runs below the electroweak scale as well. Omitting this running spoils the cancellation and produces the running of the mixing found in \Refe~\cite{Babu:2021cxe}. Figure~\ref{fig:running_examples}C confirms this numerically.

\smallskip

\textit{Model~1.---}The results that \Refe~\cite{Babu:2021cxe} plots are for its Model~1, a scotogenic model. Reference~\cite{Mir:2025fae} uses the same model to predict an excess of $\nu_\tau$-induced double-bang events in IceCube. In Model~1, the light-neutrino masses arise from a loop with the sterile neutrinos, $N$, and an inert scalar doublet, \ie, a second scalar doublet without a vacuum expectation value. The inert doublet, of mass $M_H$ at or above the electroweak scale, is the only link between $N$ and the neutrinos $\nu_\alpha$. Below $M_H$, it is integrated out, so $N$ and $\varphi$ can no longer change the neutrino mass matrix, up to corrections suppressed by powers of $M_k/M_H$. Hence, the mass matrix is generated at $\mu \sim M_H$ and does not run with $Y_N$ below that scale. Moreover, in the amplitude, $Y_N$ enters only through the self-energy at $p^2 \simeq 0$ and through vertex diagrams with the inert scalars; the dependence of the latter on $Q$ is suppressed by $Q^2/M_H^2$. Therefore, the running that \Refes~\cite{Babu:2021cxe, Mir:2025fae} find in Model~1 is absent as well, as is the excess of double bangs predicted in \Refe~\cite{Mir:2025fae}.

\smallskip

\textit{In short:} the running claimed in \Refes~\cite{Babu:2021cxe, Mir:2025fae} is absent even from the parameters of the Lagrangian. In Model~2, it cancels once the Dirac masses run; in Model~1, the mass matrix does not run with $Y_N$ below $M_H$.

\subsection{Standard Model effective field theory}
\label{sec:examples-smeft}

Reference~\cite{Bustamante:2026zst} computed the running of the mixing between production and detection in the SM, the MSSM, and the SM effective field theory (SMEFT)~\cite{Buchmuller:1985jz, Grzadkowski:2010es}. The SMEFT extends the SM by operators of dimension higher than four, suppressed by powers of the scale of the new physics, $\Lambda$. In it, dimension-six operators such as $O_{H\ell}^{(3)} = (H^\dagger i\overleftrightarrow{D}_\mu^I H)(\overline{\ell} \tau^I \gamma^\mu \ell)$ modify \equ{rge}; here, $H$ is the Higgs doublet, $\ell$ is the lepton doublet, $\tau^I$ are the Pauli matrices, and $\overleftrightarrow{D}_\mu^I$ is the Hermitian covariant derivative of \Refe~\cite{Grzadkowski:2010es}.

In the SM, the MSSM, and the SMEFT alike, \Refe~\cite{Bustamante:2026zst} took the mixing at production to be the measured one and the mixing at detection to be the running mixing matrix at the momentum transfer of detection, $U(\mu = Q_d)$. The step that fails is not the running, but the identification of $U(Q_d)$ with the mixing seen at detection, \ie, the premise. In the SMEFT, the modification of the running by $O_{H\ell}^{(3)}$ comes from loops that correct the neutrino self-energy. With $\mu = Q_d$, the self-energy, including these loops, contains large logarithms that rotate $U(Q_d)$ back to $\Uhat$ (Sec.~\ref{sec:general-muQ}). That work kept $U(Q_d)$ but dropped these logarithms. Once they are kept, detection at every $Q$ sees the physical mixing, $\Uhat$, as in Sec.~\ref{sec:general}.

The leading physical effect of $O_{H\ell}^{(3)}$ on charged-current interactions is not a change of the running, but a tree-level shift of the $W$ couplings to leptons, $\delta g_{\alpha\beta} \propto v^2 C_{H\ell,\alpha\beta}^{(3)}/\Lambda^2$, where $C^{(3)}_{H\ell}$ is the Wilson coefficient of the operator. The shift does not depend on $Q$, but it differs between flavors. The vertex then carries $(1 + \delta g)\Uhat$, which is not unitary, instead of $\Uhat$. The off-diagonal entries of $\delta g$ also mediate charged-lepton flavor violation at tree level. Hence, the shift is constrained by tests of lepton-flavor universality, of the unitarity of the mixing, and of charged-lepton flavor violation~\cite{Fernandez-Martinez:2016lgt, ParticleDataGroup:2024cfk}, not by the $Q$-dependence of the flavor composition. Four-fermion operators that couple leptons to quarks do produce effects that grow with $Q$ at the detection vertex; these are new interactions at production and detection, not a running mixing (Sec.~\ref{sec:legit}).

\smallskip

\textit{In short:} in the SMEFT, the physical effect of $O_{H\ell}^{(3)}$, a dimension-six operator that couples the Higgs doublet to the lepton doublet, is a tree-level change of the charged current that does not depend on $Q$, not a running mixing.

\section{What existing data already say}
\label{sec:data}

\subsection{Zero-baseline and short-baseline bounds}

The premise has a consequence that can be tested directly, independently of any theory argument. If the mixing at detection differed from the mixing at production, a beam of $\nu_\mu$ from pion decay, produced at $Q_p \simeq m_\pi$, would show flavor change even at zero distance from the source (Sec.~\ref{sec:general-consequences}). In the settings we consider, the mixing changes only above $Q_p \simeq m_\pi$, so the mixing at production is $\Uhat$. The transition probability is then~\cite{Babu:2021cxe, Ge:2023azz}
\begin{equation}
 \label{equ:zero_baseline}
 P_{\alpha\beta}(L = 0)
 =
 \Big| \sum_i \Uhat_{\alpha i}^\ast \, U_{\beta i}(Q_d) \Big|^2 .
\end{equation}
Short-baseline experiments that detect neutrinos via deep-inelastic scattering, at $Q_d$ of a few GeV, have searched for such transitions. They analyzed their data as two-flavor oscillations with amplitude $\sin^2 2\vartheta$ and mass splitting $\Delta m^2$. At large $\Delta m^2$, the oscillations average out over the source-detector distance. The transition probability is then $P = \frac{1}{2}\sin^2 2\vartheta$, independent of distance, as for a zero-baseline effect. NuTeV found $\sin^2 2\vartheta < 1.6 \times 10^{-3}$ and $1.1 \times 10^{-3}$, \ie, $P_{\mu e} < 8.0 \times 10^{-4}$ and $P_{\bar\mu\bar e} < 5.5 \times 10^{-4}$~\cite{NuTeV:2002daf}. NOMAD found $P_{\mu\tau} < 1.63 \times 10^{-4}$~\cite{NOMAD:2001xxt}. All limits are at 90\%~C.L. Under the premise, $P_{\alpha\beta}(0)$ is the same for neutrinos and antineutrinos, so both NuTeV limits apply.

\subsection{Model-independent bounds}

Under the premise, which Sec.~\ref{sec:general} shows to be incorrect, the zero-baseline probability, \equ{zero_baseline}, depends only on the mixing at production, $\Uhat$, and at detection, $U(Q_d)$, not on the mechanism that makes them differ. Hence, the NuTeV and NOMAD limits constrain the difference between $U(Q_d)$ and $\Uhat$, whatever the mechanism.

To translate these limits into bounds on the mixing parameters, we shift one of $\theta_{12}$, $\theta_{23}$, $\theta_{13}$, and $\delta_{\rm CP}$ at a time in $U(Q_d)$ and keep the other parameters as in $\Uhat$, including the Majorana phases, which then cancel. The limits on $P_{\mu e}$, the weaker of the two NuTeV limits, and on $P_{\mu\tau}$ then require $|\Delta\theta_{23}| < 0.7^\circ$, $|\Delta\theta_{12}| < 2.3^\circ$, $|\Delta\theta_{13}| < 2.4^\circ$, and $|\Delta\delta_{\rm CP}| < 16^\circ$ between $Q \simeq m_\pi$ and $Q$ of a few GeV. Therefore, if the premise held, any model that changes only one of these parameters by more than these amounts between these scales would already be excluded. Simultaneous changes of several parameters could partly compensate one another.

These bounds apply to the models of \Refe~\cite{Babu:2021cxe}, in which new light particles would make the mixing run below the electroweak scale; to the configuration of \Refe~\cite{Mir:2025fae} in which the running starts at $m_\pi$; to the model-independent parametrizations of \Refes~\cite{Ge:2023azz, Denton:2026mkp}, when applied between pion decay and detection at momentum transfers up to a few GeV; and to the HNL example (Sec.~\ref{sec:examples-hnl}). References~\cite{Babu:2021cxe, Babu:2022non, Ge:2023azz} already imposed zero-baseline limits from NuTeV, NOMAD, CHARM-II, ICARUS, or CDHS. In \Refe~\cite{Babu:2021cxe}, these limits restrict the effects at T2K and NOvA to perturbations. In \Refe~\cite{Babu:2022non}, they apply to the running mixing with sterile neutrinos.

The bounds do not apply when the running starts above the momentum transfers of these experiments. This is the case in the MSSM example (Sec.~\ref{sec:examples-mssm}). It is also the case in the configuration of \Refe~\cite{Mir:2025fae} that produces its double-bang signal, in which the new states have masses near 5~GeV. There, only the theory arguments apply: that of Sec.~\ref{sec:general} and, for \Refe~\cite{Mir:2025fae}, that of Sec.~\ref{sec:examples-model2}.

In the parametrization of \Refes~\cite{Ge:2023azz, Ge:2024ibn, Denton:2026mkp}, each mixing parameter, $X$, runs as $dX/d\ln|Q^2| = \beta_X$, with $\beta_X$ constant, so that $\Delta X = \beta_X \ln(Q_d^2/m_\pi^2)$ between pion decay and detection. Since $\ln(Q_d^2/m_\pi^2) \approx 5$--$7$ for $Q_d$ of 2--5~GeV, these bounds imply $|\beta_{12}|, |\beta_{13}| \lesssim 0.006$--$0.008$, $|\beta_{23}| \lesssim 0.002$, and $|\beta_\delta| \lesssim 0.04$--$0.05$. The bound on $\beta_\delta$ agrees with that of \Refe~\cite{Ge:2023azz} and excludes the value $\beta_\delta = 0.15$ that \Refe~\cite{Ge:2024ibn} uses to illustrate its reactor signal. The bounds on the angles are tighter than the joint sensitivities projected for future near detectors in \Refe~\cite{Denton:2026mkp}, $|\beta_{12}|, |\beta_{13}| \lesssim 0.015$ and $|\beta_{23}| \lesssim 0.01$ at 68\%~C.L., although that work also addresses correlations among several $\beta_X$.

\subsection{The HNL example}
\label{sec:data-hnl}

\begin{figure}[t!]
 \centering
 \includegraphics[width=\columnwidth]{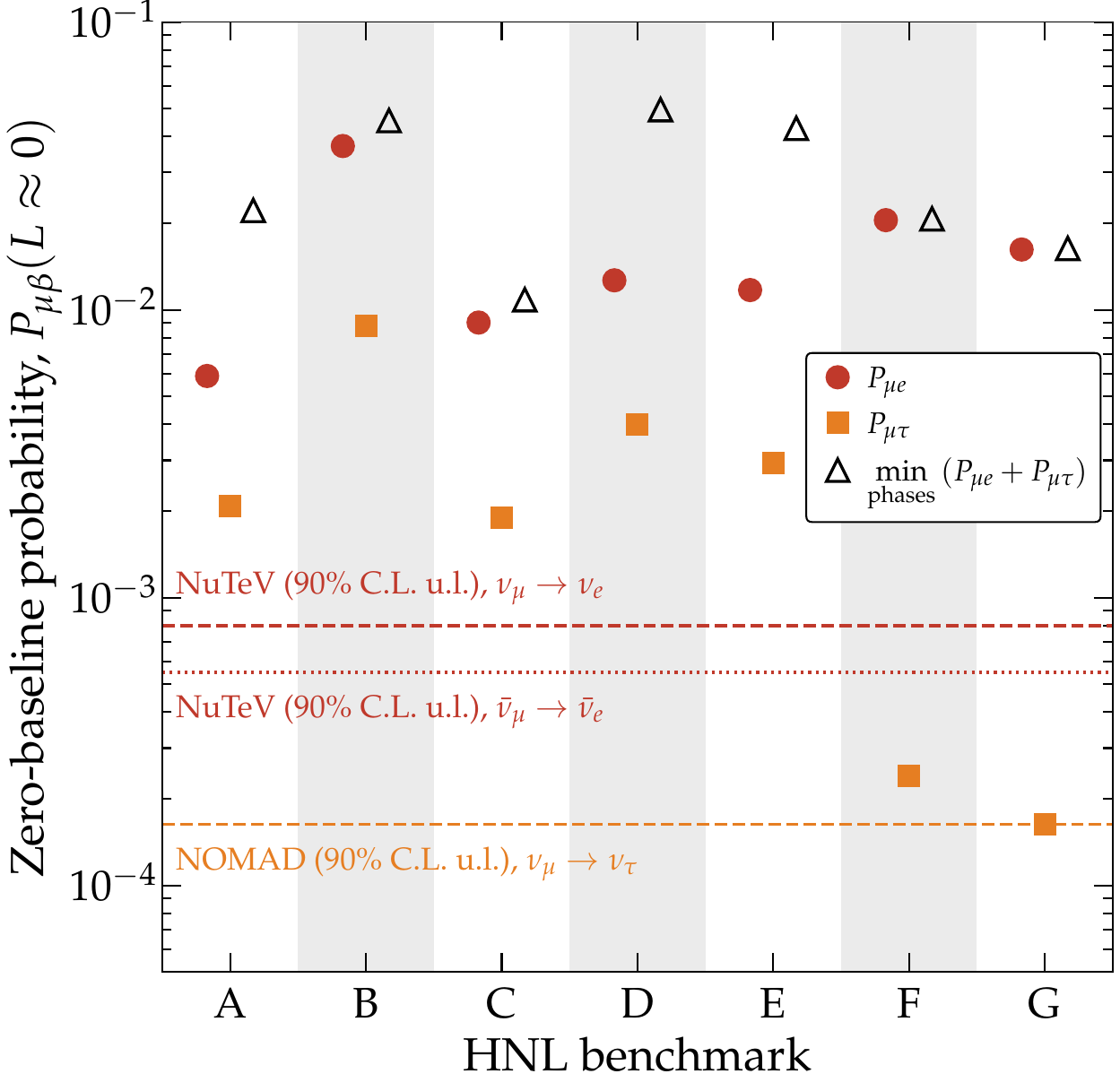}
 \caption{\textbf{\textit{Zero-baseline flavor change implied by the premise, compared with data.}} Transition probabilities at zero baseline that the premise predicts for the HNL benchmarks of Table~\ref{tab:hnl_benchmarks}, with production governed by $\Uhat$ and detection at $Q_d > M$ by $U_{\rm tree}$. The probabilities $P_{\mu e}$ and $P_{\mu\tau}$ use the Majorana phases of each benchmark (Sec.~\ref{sec:data-hnl}); the floor on $P_{\mu e} + P_{\mu\tau}$, \equ{floor}, holds for any phases of the columns of $U_{\rm tree}$. The 90\%~C.L.~upper limits are from NuTeV~\cite{NuTeV:2002daf} and NOMAD~\cite{NOMAD:2001xxt}. The physical prediction is zero.}
 \label{fig:zero_baseline}
\end{figure}

\begin{table*}[t!]
 \footnotesize
 \caption{\label{tab:hnl_benchmarks}\textbf{\textit{HNL benchmarks and what the premise would imply.}} Benchmarks A--G of \Refe~\cite{Barenboim:2026mvn}, with normal ordering, $m_1 = 0.01$~eV, NuFIT~6.1 mixing parameters~\cite{Esteban:2024eli, NuFIT61}, real couplings $U_{\alpha N}$, and Majorana phases $\rho$, $\sigma$ (multiplying the first and second columns of $\Uhat$). Columns: $|\delta m|$, the size of the single nonzero element of $\delta m$, \equ{delta_m_hnl}, for the flavor to which $N$ couples; $\max|\Delta|U|^2|$, the largest difference between $|U_{{\rm tree},\alpha i}|^2$ and $|\Uhat_{\alpha i}|^2$; $\theta_{12}^{\rm tree}$, the solar angle of $U_{\rm tree}$; and the zero-baseline transition probabilities that the premise predicts for detection at $Q > M$ (Sec.~\ref{sec:data}), the last one a lower bound for any phases of the columns of $U_{\rm tree}$. The probabilities $P_{\mu e}(0)$ and $P_{\mu\tau}(0)$ are each minimized separately over the signs of the columns of $U_{\rm tree}$, so their sum can lie below that bound. Reference~\cite{Barenboim:2026mvn} used an incorrect $\delta m$, with a spurious $W$-boson term and a $Z$-boson term of the wrong sign and coefficient, and chose its phases with it; we use \equ{delta_m_hnl}, so the shifts that the premise implies differ from those reported there. With either $\delta m$, the premise fails: the physical predictions are $P_{\mu e}(0) = P_{\mu\tau}(0) = 0$ (Sec.~\ref{sec:examples-hnl}).}
 \begin{ruledtabular}
  \begin{tabular}{c c c c c c c c c c c}
   & $M$ [GeV] & Coupling & $\rho$ [$^\circ$] & $\sigma$ [$^\circ$] & $|\delta m|$ [meV] & $\max|\Delta|U|^2|$ & $\theta_{12}^{\rm tree}$ [$^\circ$] & $P_{\mu e}(0)$ & $P_{\mu\tau}(0)$ & \begin{tabular}{@{}c@{}}$P_{\mu e}(0) + P_{\mu\tau}(0)$,\\ any phases\end{tabular} \\
   \hline
   A & 3 & $|U_{eN}|^2 = 5 \times 10^{-8}$ & 0.00 & 16.24 & 3.95 & 0.21 & 18.4 & $5.9 \times 10^{-3}$ & $2.1 \times 10^{-3}$ & $\geq 2.2 \times 10^{-2}$ \\
   B & 3 & $|U_{\mu N}|^2 = 4 \times 10^{-8}$ & 5.38 & 182.34 & 3.16 & 0.27 & 49.8 & $3.7 \times 10^{-2}$ & $8.8 \times 10^{-3}$ & $\geq 4.5 \times 10^{-2}$ \\
   C & 4 & $|U_{\mu N}|^2 = 1 \times 10^{-8}$ & 5.14 & 181.26 & 1.72 & 0.13 & 41.5 & $9.0 \times 10^{-3}$ & $1.9 \times 10^{-3}$ & $\geq 1.1 \times 10^{-2}$ \\
   D & 1 & $|U_{\tau N}|^2 = 1.5 \times 10^{-6}$ & 87.16 & 92.77 & 5.76 & 0.24 & 14.9 & $1.3 \times 10^{-2}$ & $4.0 \times 10^{-3}$ & $\geq 5.0 \times 10^{-2}$ \\
   E & 3 & $|U_{\tau N}|^2 = 6 \times 10^{-8}$ & 90.61 & 91.23 & 4.73 & 0.22 & 17.3 & $1.2 \times 10^{-2}$ & $2.9 \times 10^{-3}$ & $\geq 4.3 \times 10^{-2}$ \\
   F & 1 & $|U_{\tau N}|^2 = 7 \times 10^{-7}$ & 179.56 & 173.58 & 2.69 & 0.20 & 45.5 & $2.0 \times 10^{-2}$ & $2.4 \times 10^{-4}$ & $\geq 2.1 \times 10^{-2}$ \\
   G & 3 & $|U_{\tau N}|^2 = 3 \times 10^{-8}$ & 179.74 & 174.05 & 2.37 & 0.17 & 44.1 & $1.6 \times 10^{-2}$ & $1.6 \times 10^{-4}$ & $\geq 1.6 \times 10^{-2}$ \\
  \end{tabular}
 \end{ruledtabular}
\end{table*}

Figure~\ref{fig:zero_baseline} and Table~\ref{tab:hnl_benchmarks} show the zero-baseline transition probabilities that the premise predicts for the HNL benchmarks of Sec.~\ref{sec:examples-hnl}. Under the premise, production, at $Q_p \simeq m_\pi < M$, uses the physical mixing, $\Uhat$, while detection at $Q_d > M$ uses the mixing without $\delta m$, $U_{\rm tree}$. The two differ by far more than the model-independent bounds allow: the solar angle of $U_{\rm tree}$, $\theta_{12}^{\rm tree}$, ranges from $15^\circ$ to $50^\circ$, against the physical $33.8^\circ$. As a result, the premise predicts $P_{\mu e}(0)$ of at least $0.6\%$--$3.7\%$, \ie, 7--46 times the limit on $P_{\mu e}$, and 11--67 times the limit on $P_{\bar\mu\bar e}$.

These probabilities depend on the Majorana phases, which oscillation data do not fix. In standard oscillations, the same mixing matrix enters production and detection, so its Majorana phases cancel. Here, two different matrices enter, $\Uhat$ and $U_{\rm tree}$, so the Majorana phases do not cancel. We use the Majorana phases of each benchmark, $\rho$ and $\sigma$ (Table~\ref{tab:hnl_benchmarks}), in $\Uhat$. They fix the phases of $U_{\rm tree}$ through the requirement that the tree-level masses be real and positive. This requirement leaves the sign of each column of $U_{\rm tree}$ free, since flipping it does not change the masses. The premise does not say which sign to take. Since we test whether data exclude the premise, we take, for each probability, the signs that make it smallest: if even this smallest value exceeds the experimental limits, the premise is excluded for every choice of signs.

The requirement of real, positive masses fixes the phases of the columns of each matrix with respect to its own mass matrix, $\meff$ for $\Uhat$ and $m_{\rm tree}$ for $U_{\rm tree}$. The premise, however, uses both matrices for the same propagating states without relating the phases of their columns. To show that the conclusion does not depend on this relation, we also compute the smallest value of the total appearance probability, $P_{\mu e} + P_{\mu\tau}$, over arbitrary phases of the columns of $U_{\rm tree}$. Under the premise, $P_{\mu e} + P_{\mu\tau} = 1 - P_{\mu\mu}$, with $P_{\mu\mu} = |\sum_i \Uhat_{\mu i}^\ast U_{{\rm tree},\mu i}|^2$. The modulus of this sum is largest when its three terms have the same phase, so that
\begin{equation}
 \label{equ:floor}
 P_{\mu e} + P_{\mu\tau}
 \geq
 1 - \Big( \sum_i |\Uhat_{\mu i}| \, |U_{{\rm tree},\mu i}| \Big)^2 .
\end{equation}
The right-hand side depends only on the moduli of the matrix elements, which are fixed for each benchmark. It ranges from $1.1\%$ to $5.0\%$, more than ten times the sum of the limits on $P_{\mu e}$ and $P_{\mu\tau}$, $9.6 \times 10^{-4}$. Hence, the premise would be excluded even if only one in ten deep-inelastic events in these experiments had $Q_d > M$.

\subsection{High-energy astrophysical neutrinos}

High-energy astrophysical neutrinos are thought to come from the decay of charged pions made in interactions of high-energy protons inside astrophysical sources. The full decay chain, $\pi^+ \to \mu^+ \nu_\mu$ followed by $\mu^+ \to e^+ \nu_e \bar\nu_\mu$, and its charge conjugate yield the flavor composition at the sources, $(1{:}2{:}0)_{\rm S}$, \ie, $f_{e,{\rm S}} = 1/3$, $f_{\mu,{\rm S}} = 2/3$, and $f_{\tau,{\rm S}} = 0$. If the muons lose energy before decaying, \eg, in strong magnetic fields, the composition is $(0{:}1{:}0)_{\rm S}$. Over cosmological distances, the oscillations average out, so the flavor composition at Earth depends only on the mixing (Sec.~\ref{sec:general-consequences}); see, \eg, \Refe~\cite{Bustamante:2015waa}.

\begin{figure*}[t!]
 \centering
 \includegraphics[width=0.92\textwidth]{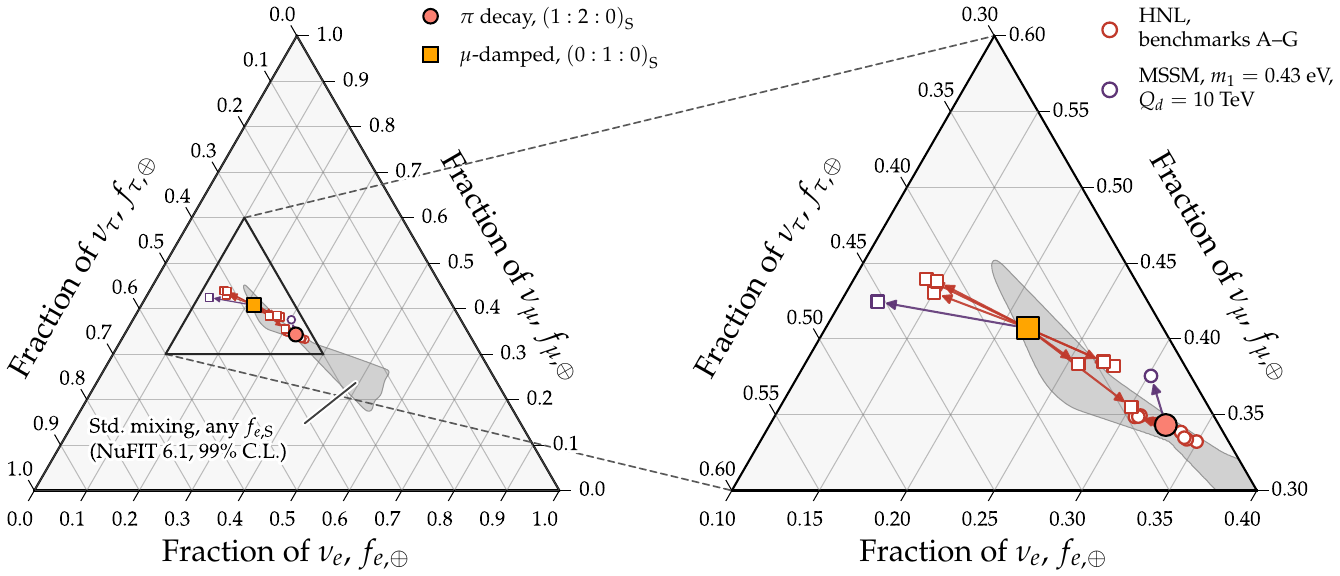}
 \caption{\textbf{\textit{Flavor composition of high-energy astrophysical neutrinos at Earth.}} The physical predictions use $\Uhat$, with the NuFIT~6.1 best-fit mixing parameters for normal ordering, for neutrinos produced by full pion decay, $(1{:}2{:}0)_{\rm S}$, or by pion decay with muons that lose energy before decaying, $(0{:}1{:}0)_{\rm S}$; they hold at every momentum transfer. The standard-mixing region spans all source compositions with $f_{\tau,{\rm S}} = 0$, for the mixing parameters allowed at 99\%~C.L.~\cite{Bustamante:2015waa, Song:2020nfh}. Under the premise, detection at momentum transfer $Q_d$ would use $U(Q_d)$ instead of $\Uhat$, which would shift the flavor composition: for the HNL benchmarks A--G (Table~\ref{tab:hnl_benchmarks}), through the step at $Q = M$; for the MSSM example with $m_1 = 0.43$~eV, through the running up to $Q_d = 10$~TeV. The mixing does not change with $Q$, so none of these shifts occurs.}
 \label{fig:flavor_triangle}
\end{figure*}

Figure~\ref{fig:flavor_triangle} shows the shifts in this composition that the premise predicts for the HNL benchmarks and for the MSSM example. Under the premise, detection uses $U(Q_d)$ instead of $\Uhat$, so the flavor composition at Earth would be $f_{\beta,\oplus} = \sum_\alpha f_{\alpha,{\rm S}} \sum_i |\Uhat_{\alpha i}|^2 |U_{\beta i}(Q_d)|^2$, instead of the expression of Sec.~\ref{sec:general-consequences}, which has $\Uhat$ in both places. For the HNL benchmarks, the flavor fractions would shift by up to 0.023 for $(1{:}2{:}0)_{\rm S}$ and by up to 0.085 for $(0{:}1{:}0)_{\rm S}$; for the MSSM example, by up to 0.032 for $(1{:}2{:}0)_{\rm S}$ and by up to 0.094 for $(0{:}1{:}0)_{\rm S}$. Present measurements of the flavor composition cannot yet test these shifts, because they do not constrain the mixing parameters~\cite{Bustamante:2026aur}. Nevertheless, since the premise fails, none of these shifts occurs. Hence, a flavor composition measured outside the standard-mixing region of \figu{flavor_triangle} would point to other physics (Sec.~\ref{sec:legit}).

\section{Common misconceptions}
\label{sec:misconceptions}

Several arguments make the premise appear natural. We list them below, with the reason each one fails.

\begin{enumerate}[leftmargin=*]
 \item
  \textit{``High-energy experiments probe high $\mu$.''---}The renormalization scale is a bookkeeping device of the calculation, not a property of an experiment (Sec.~\ref{sec:general-running}). It is chosen to keep small the logarithms of ratios of the scales that appear in a given loop. For the mixing, the relevant loop is the neutrino self-energy, whose scales are the masses of the particles in it, not the energy or the momentum transfer of the process (Sec.~\ref{sec:general-physical}).

 \item
  \textit{``Vertex corrections are evaluated at $\mu = Q$, so the mixing in the vertex is $U(Q)$.''---}Choosing $\mu = Q$ is legitimate, but then the self-energy on the external neutrino leg contains large logarithms, $\ln(Q^2/M_X^2)$, that rotate $U(Q)$ back to $\Uhat$ (Sec.~\ref{sec:general-muQ}). The mixing is not a coupling of the vertex. In the flavor basis, the $W$ couples with equal strength to the three flavor states; the mixing matrix appears at the vertex only when the neutrino is written in terms of mass eigenstates [\equ{cc}]. It describes the flavor content of the external neutrino, which is fixed by the neutrino propagator, including its self-energy (Sec.~\ref{sec:general-physical}).

 \item
  \textit{``Running couplings and masses are observed, \eg, $\alpha_s(Q)$ and $m_b(Q)$; so running mixing should be observable too.''---}A quantity measured in a process can depend on $Q$ only if the quantum corrections that determine it involve $Q$. The strong coupling measured in a QCD process is determined by corrections at the interaction point, \ie, gluon loops and emissions, whose momenta grow with $Q$. Hence, it depends on $Q$: this is the observed running of $\alpha_s(Q)$. The same holds for the bottom-quark mass extracted from $e^+ + e^- \to b + \bar b + g$ at the $Z$ pole, $m_b(M_Z)$~\cite{DELPHI:1997rxp, Brandenburg:1999nb}. The mixing measured in a neutrino process, instead, is determined by the self-energy, \ie, the correction to the neutrino propagator, which does not involve $Q$, because the neutrino stays on its mass shell and $Q$ is carried by the $W$ (Sec.~\ref{sec:general-physical}). The corrections at the interaction point, which do involve $Q$, are small for the mixing (Sec.~\ref{sec:general-vertex}). The weak coupling does run with $Q$ in the same process, through the self-energy of the $W$ and the vertex corrections, but this running is the same for every flavor, so it does not change the mixing (Sec.~\ref{sec:general-amplitude}). Hence, the physical mixing, $\Uhat$, does not depend on $Q$. Like the pole mass of a freely propagating particle, it is fixed by the propagator. The running mixing matrix, $U(\mu)$, is a parameter of the Lagrangian, like the running mass, $m_b(\mu)$.

 \item
  \textit{``Propagation uses pole masses, but production and detection use running mixing.''---}This is the explicit setup of \Refes~\cite{Babu:2021cxe, Babu:2022non, Ge:2023azz}. It is implicit in the other works in Table~\ref{tab:literature}, which take the propagating states to be the physical mass eigenstates. However, masses and mixing are the eigenvalues and eigenvectors of one matrix, $\meff$ (Sec.~\ref{sec:general-physical}). The propagating states are its eigenvectors, so the vertex couples to them through $\Uhat$. Taking the eigenvalues from $\meff$ and the eigenvectors from $m(Q)$ is not consistent (Sec.~\ref{sec:general-muQ}).

 \item
  \textit{``A hard probe resolves the heavy particles that generate the correction.''---}A probe resolves the structure of the line through which its momentum flows. The momentum transfer is carried by the $W$; the neutrino stays on its mass shell (Sec.~\ref{sec:general-amplitude}). Heavy states are resolved by virtual neutrino lines with $|p^2| \gtrsim M^2$ (Sec.~\ref{sec:examples-hnl}), not by the momentum transfer of the $W$. In the same way, a hard collision does not change the $K^0$--$\bar K^0$ composition of an incoming $K_L$ (Sec.~\ref{sec:synopsis}).

 \item
  \textit{``The mixing changes at thresholds.''---}Step functions at thresholds are an artifact of mass-independent schemes, such as $\overline{\rm MS}$, in which heavy particles are integrated out abruptly at their mass. Observables vary smoothly. In the HNL example, even $U(\mu = Q)$ would change gradually, logarithmically, between $M$ and $M_h$ (\figu{running_examples}B). The physical mixing does not change at all: it is $\Uhat$ at every $Q$ (Sec.~\ref{sec:examples-hnl}).

 \item
  \textit{``Flavor eigenstates are energy-dependent.''---}This impression has two sources. First, the premise identifies $\mu$ with $Q$, which typically grows with the neutrino energy in deep-inelastic scattering, so the running mixing is often shown as a function of energy (Sec.~\ref{sec:synopsis}). Second, in matter, the eigenstates of the Hamiltonian that governs propagation, and hence the effective mixing, do depend on energy~\cite{Wolfenstein:1977ue, Mikheyev:1985zog}. Neither changes the flavor states. A flavor state is the state produced or detected with a given charged lepton. In terms of the physical mass eigenstates, it is $\sum_i \Uhat_{\alpha i}^\ast |\hat\nu_i\rangle$, up to the small corrections in \equ{amplitude}, at any energy and momentum transfer (Sec.~\ref{sec:general-consequences}). Matter effects act during propagation, through a Hamiltonian with the same $\Uhat$.

 \item
  \textit{``A neutrino of higher energy probes the mixing at a higher scale.''---}The energy of a neutrino cannot set the scale of its mixing, because the energy depends on the reference frame (Table~\ref{tab:scales}): a Lorentz boost turns a PeV neutrino into an MeV one, or brings it to rest. The mixing, a property of the neutrino, cannot depend on the frame. Indeed, the self-energy that fixes it depends on the neutrino four-momentum only through the frame-independent $p^2 = m_i^2$; the momentum transfer, $Q$, which is also frame-independent, does not enter it (Sec.~\ref{sec:general-amplitude}).

 \item
  \textit{``Then the RG running of neutrino parameters is useless.''---}It is not. It is how a model defined at a high mass scale, such as that of grand unification, is confronted with the measured mixing~\cite{Antusch:2001ck, Antusch:2003kp, Antusch:2005gp}. In the example of Sec.~\ref{sec:antusch-use} (\figu{rg_use}), a model that fixes its parameters at $10^{16}$~GeV, with $\tan\beta = 50$ and $m_1 = 0.01$~eV, must set $\theta_{12}(10^{16}~{\rm GeV}) = 29.2^\circ$ to produce the best-fit measured angle, $33.8^\circ$. Thus, the running is essential for testing flavor symmetries, grand unification, and leptogenesis. What it does not do is make different experiments see different mixing.

 \item
  \textit{``Relating a high-scale prediction to low-energy data is the same as relating production to detection.''---}It is not. In the first case, the running converts the input of a model, fixed at the scale where the model generates the flavor structure, into the physical mixing, $\Uhat$, to which $U(\mu)$ reduces for $\mu \lesssim M_X$ (Sec.~\ref{sec:antusch-same}). In the second case, the running would have to convert one physical mixing, seen at $Q_p$, into another, seen at $Q_d$. There is only one physical mixing. The momentum transfer of a process is not a scale of the theory and does not appear in the neutrino self-energy (Sec.~\ref{sec:general-physical}). The physical mixing is the same at production and detection even for a model whose parameters are fixed at $\mu \simeq Q_p$ (Sec.~\ref{sec:antusch-use}).
\end{enumerate}

\section{What can make the measured flavor differ from expectations}
\label{sec:legit}

Our result does not imply that the flavor measured in an experiment, terrestrial or astrophysical, must agree with the standard expectation, computed with $\Uhat$ (Sec.~\ref{sec:general-consequences}). It implies that the RG running of the mixing is not a reason for it to disagree. Legitimate reasons, reviewed in \Refe~\cite{Arguelles:2022tki}, fall into three classes:
\begin{enumerate}[leftmargin=*]
 \item
  \textit{New physics in propagation.---}New interactions with matter~\cite{Wolfenstein:1977ue, Farzan:2017xzy, Proceedings:2019qno} and Lorentz- or CPT-violating terms~\cite{Kostelecky:2003cr} add a term, $V(E)$, to the Hamiltonian that governs propagation, $H = \Uhat \, {\rm diag}(\hat m_i^2) \Uhat^\dagger/2E + V(E)$. Scattering on dark matter~\cite{Arguelles:2017atb} or on the cosmic neutrino background~\cite{Ng:2014pca, Bustamante:2020mep, Esteban:2021tub, Berryman:2022hds}, neutrino decay~\cite{Beacom:2002vi}, and decoherence~\cite{Hooper:2004xr} also modify propagation. Thus, all of these effects change the flavor composition of high-energy astrophysical neutrinos at Earth, which travel the longest distances at the highest energies~\cite{Bustamante:2010nq, Arguelles:2015dca, Bustamante:2015waa, Shoemaker:2015qul, Rasmussen:2017ert, Ahlers:2018mkf, Ackermann:2022rqc}. When $V$ grows with energy, the eigenstates of $H$, and hence the effective mixing in propagation, change with energy~\cite{Arguelles:2015dca, Rasmussen:2017ert, Ahlers:2018yom, Ahlers:2020miq, IceCube:2021tdn}. This energy dependence does not contradict our result. New physics in propagation changes how the flavor states evolve between production and detection, not the flavor states themselves, whose composition in terms of the mass eigenstates is given by $\Uhat$ at any energy and momentum transfer (Sec.~\ref{sec:misconceptions}).

  Measurements of the energy dependence of the flavor composition test these effects~\cite{Liu:2023flr}, as will measurements of the mixing parameters above 1~TeV, once neutrino telescopes constrain them~\cite{Bustamante:2026aur}. The analyses of \Refes~\cite{Liu:2023flr, Bustamante:2026aur} do not rely on the premise: both use a single mixing matrix at production and detection.
  
 \item
  \textit{New interactions at production and detection.---}Contact interactions or light mediators that couple leptons to quarks modify the production and detection amplitudes through flavor-dependent form factors that can depend on $Q$~\cite{Rasmussen:2017ert}. They enter these amplitudes as $\mathcal{A}_{\beta i} \propto [(1 + \eta(Q^2)) \Uhat]_{\beta i}$, like $\Gamma$ in \equ{amplitude} but without its loop suppression; the matrix $1 + \eta$ is generally not unitary. Such effects should be parametrized as $Q$-dependent, generally non-unitary corrections, not as unitary mixing matrices that differ between production and detection. They are subject to bounds from lepton-flavor universality and from searches for non-standard interactions, which limit them to below about $10^{-2}$--$10^{-1}$~\cite{Grossman:1995wx, Biggio:2009nt, Farzan:2017xzy, Proceedings:2019qno} (Sec.~\ref{sec:general-consequences}).

 \item
  \textit{Non-unitarity.---}Mixing with heavy states makes the $3 \times 3$ mixing matrix non-unitary, at $\mathcal{O}(|U_{\alpha N}|^2)$, independently of $Q$~\cite{Antusch:2006vwa}; it has the same form as the $Q$-independent correction $\Delta$ of Sec.~\ref{sec:general-consequences}. It changes the flavor transitions expected in every experiment, at all energies~\cite{Antusch:2006vwa, Ahlers:2020miq}, by amounts that data bound to below about $10^{-3}$~\cite{Fernandez-Martinez:2016lgt}.
\end{enumerate}

Phenomenological parametrizations in which production and detection use different unitary mixing matrices~\cite{Bustamante:2026zst, Ge:2023azz, Ge:2024ibn, Ge:2026hsh, Denton:2026mkp} correspond, at most, to the special case of the second class in which $\eta$ is anti-Hermitian, so that $1 + \eta$ is a rotation. The analyses that use them could be reinterpreted as tests of new interactions at production and detection if the parametrization is generalized to a non-unitary $1 + \eta$.

\section{Conclusions}
\label{sec:conclusions}

The renormalization-group (RG) running of the neutrino mixing parameters is real and useful, but it is undetectable: the mixing does not depend on the momentum transfer of the process that produces or detects the neutrino. A growing body of work (Table~\ref{tab:literature}) assumes the opposite. If this premise held, experiments at different momentum transfers would measure different mixing parameters; neutrinos would even change flavor at zero distance from their source. We have shown where and why the premise fails, so that future work can avoid it.

The RG running concerns $U(\mu)$, the mixing matrix as a parameter of the Lagrangian, defined at an arbitrary renormalization scale, $\mu$. The premise takes the mixing of a neutrino produced or detected in a process with momentum transfer $Q$ to be $U(\mu = Q)$. It thereby conflates $U(\mu)$ with the mixing that experiments measure, $\Uhat$, a property of the neutrino itself. Because the neutrino travels a macroscopic distance between production and detection, it is on its mass shell, $p^2 = m_i^2$, where its propagator, including its self-energy, fixes the mixing. The self-energy depends on the masses of the particles in its loop, not on $Q$, which flows through the $W$, not through the neutrino.

The premise borrows the choice $\mu = Q$ from running couplings, such as the strong coupling, for which it keeps the logarithms in the corrections to a process small. For the mixing, the choice is legitimate but has the opposite effect: the production and detection amplitudes then contain $U(Q)$, while the self-energy acquires large logarithms of the ratio of $Q$ to the masses in its loop, which rotate $U(Q)$ back to $\Uhat$. The premise keeps $U(Q)$ but incorrectly drops these logarithms. Once we keep them, the only dependence on $Q$ that can affect the flavor comes from the vertex and box corrections at production and detection; loop factors and couplings suppress these corrections. \textbf{Hence, every experiment, at any momentum transfer, measures the same mixing, $\Uhat$.}

We have shown this explicitly in four examples: the MSSM at large $\tan\beta$, a seesaw with a GeV-scale heavy neutral lepton (HNL), a seesaw with a light scalar, and the SMEFT. Independently of theory, short-baseline data already limit the change of any one mixing angle between neutrino production in pion decay and detection at momentum transfers of a few GeV to less than $0.7^\circ$--$2.4^\circ$. In the HNL example, the premise predicts flavor change at zero distance from the source more than ten times larger than these data allow.

Neither RG running nor threshold corrections that switch on or off with $Q$ produce the signatures proposed under the premise, such as flavor change at zero distance from the source, shifts in the measured CP-violation phase, and an anomalous flavor composition of high-energy astrophysical neutrinos. This applies to every work in Table~\ref{tab:literature}, including our own. RG running remains the standard tool to relate models defined at a high mass scale to the measured mixing; loop corrections to the neutrino masses from heavy states still constrain low-scale seesaw models. Flavor measurements, terrestrial and astrophysical, remain tests of new physics in propagation, of new interactions at production and detection, and of non-unitarity. Analyses that look for a dependence on $Q$ at production and detection should parametrize it as a generally non-unitary correction to the amplitudes, not as a change of the mixing.

\begin{acknowledgments}

MB is supported by {\sc Villum Fonden} under project no.~29388.  QL is supported by the Canada First Research Excellence Fund and the Natural Sciences and Engineering Research Council of Canada through the Arthur B.~McDonald Canadian Astroparticle Physics Research Institute.  GB is supported by the Spanish grants CIPROM/2021/054 (Generalitat Valenciana) and PID2023-151418NB-I00, funded by MICIU/AEI/10.13039/501100011033, and by the European ITN project HIDDeN (H2020-MSCA-ITN-2019/860881-HIDDeN). During the preparation of this work, the authors used Claude for research assistance in code development, numerical auditing, figure preparation, and editorial revision. All scientific judgments, validation decisions, and responsibility for the manuscript remain with the authors.

\end{acknowledgments}

\appendix

\section{One-loop two-point function and the physical mixing}
\label{app:twopoint}

At one loop, Eqs.~(\ref{equ:meff}) and (\ref{equ:uhat}) are the standard definition of the mixing matrix in the on-shell renormalization scheme. For external fermions on their mass shell, the off-diagonal wave-function renormalization, \ie, the correction that mixes different fields, carries factors $1/(m_i^2 - m_j^2)$, which produce the degeneracy enhancement of \equ{enhancement}. The on-shell scheme absorbs this enhanced part into the definition of the physical mixing. References~\cite{Denner:1990yz, Denner:1991kt, Gambino:1998ec, Pilaftsis:2002nc, Denner:2004bm, Kniehl:2006bs, Kniehl:2006rc} do this for the quark mixing matrix; \Refes~\cite{Kniehl:1996bd, Almasy:2009kn, Huang:2025qew} do it for Majorana neutrinos, including seesaw models. In the gauge-independent formulation of \Refes~\cite{Kniehl:2006bs, Kniehl:2006rc, Almasy:2009kn}, the mixing matrix diagonalizes the mass matrix, including the mass-like parts of the self-energy, as \equ{uhat} does; the remaining corrections to the amplitude are free of the factors $1/(m_i^2 - m_j^2)$.

We sketch how \equ{meff} follows from the standard on-shell renormalization of the mixing matrix~\cite{Denner:1990yz, Kniehl:1996bd, Almasy:2009kn}. In four-component notation, with $P_{L,R} = (1 \mp \gamma_5)/2$, the one-loop self-energy of the light Majorana neutrinos is~\cite{Kniehl:1996bd}
\begin{eqnarray}
 \Sigma_{ij}(p) &=& \slashed{p} \left[ P_L \Sigma^L_{ij}(p^2) + P_R \Sigma^R_{ij}(p^2) \right] \nonumber \\
 && + \, P_L \Sigma^M_{ij}(p^2) + P_R \Sigma^{M\ast}_{ij}(p^2) \;,
\end{eqnarray}
in the basis that diagonalizes $m(\mu)$. Here, $\Sigma^L$ and $\Sigma^R$ correct the kinetic terms of the left- and right-chiral components; $\Sigma^M$ corrects the mass term. Together, they play the roles of $\Sigma_K$ and $-\Sigma_M$ in \equ{two_point}, since here $\Sigma$ adds to the inverse propagator, as in \Refe~\cite{Kniehl:1996bd}. The on-shell conditions require the renormalized propagator to have its poles at the physical masses, with unit residues and no mixing between different mass eigenstates on their mass shell. The last condition fixes the off-diagonal wave-function renormalization constants, schematically, as
\begin{equation}
 \delta Z^L_{ij}
 \propto
 \frac{m_j^2 \Sigma^L_{ij} + m_i m_j \Sigma^R_{ij} + m_i \Sigma^M_{ij} + m_j \Sigma^{M\ast}_{ij}}{m_i^2 - m_j^2} \;,
\end{equation}
evaluated at $p^2 = m_j^2$; \Refe~\cite{Kniehl:1996bd} gives the exact expressions. The anti-Hermitian part of $\delta Z^L$ is a rotation, which carries the factors $1/(m_i^2 - m_j^2)$. Combined with $U(\mu)$, the matrix of eigenvectors of $m(\mu)$, it yields $\Uhat$, the matrix of eigenvectors of $\meff$ [\equ{meff}]. In the Hermitian part of $\delta Z^L$, these factors cancel, up to corrections of order $m_i^2/M_X^2$; this part enters $\Delta$ in \equ{amplitude}. For loop masses much larger than $m_i$, the self-energies vary slowly near $p^2 \simeq 0$, so they can be evaluated at $p^2 = 0$. The construction reduces to diagonalizing $\meff$. Since the on-shell conditions involve the neutrino's own $p^2 = m_j^2$, not the momentum transfer, $\Uhat$ is the same for any process.

\section{One-loop correction from a GeV-scale HNL}
\label{app:hnl}

Equation~(\ref{equ:delta_m_hnl}) follows from the $Z$ and Higgs loops of \figu{selfenergies}(b), in agreement with \Refes~\cite{Pilaftsis:1991ug, Grimus:2002nk, Lopez-Pavon:2015cga}; in a renormalizable gauge, the $Z$ term includes the loop of the neutral Goldstone boson. Below $M_Z$, the logarithms of $\delta m$ also follow from the low-energy effective field theory of \Refe~\cite{Jenkins:2017jig}, extended by $N$. In it, four-fermion operators with two $N$ fields replace the exchange of the $Z$ and Higgs bosons; running these operators from $M_Z$ down to $M$ generates the logarithms. For $M \ll M_Z$, the bracket in \equ{delta_m_hnl} is $3\ln(M_Z^2/M^2) + \ln(M_h^2/M^2) = 4\ln(M_Z^2/M^2) + \ln(M_h^2/M_Z^2)$. The first term comes from running between $M_Z$ and $M$; the second, a small term, from the difference between $M_h$ and $M_Z$. The leading-logarithmic curve for $U(\mu = Q)$ in \figu{running_examples}B keeps only the part of each logarithm that comes from loop momenta above $Q$. For $Q > M$, the $Z$ term becomes $3\ln(M_Z^2/Q^2)$ below $M_Z$ and vanishes above it; the Higgs term becomes $\ln(M_h^2/Q^2)$ below $M_h$ and vanishes above it.

\section{Model 2: self-energy and running}
\label{app:model2}

This appendix supports Sec.~\ref{sec:examples-model2}. Reference~\cite{Babu:2021cxe} proposed two models in which light new particles make the mixing run below the electroweak scale. Its Model~2, \equ{model2_lagrangian}, is a type-I seesaw with light sterile neutrinos, $N_R$, and a light scalar singlet, $\varphi$, that carries lepton number. The Dirac masses, $m_D = Y_\nu v/\sqrt{2}$, connect the sterile neutrinos to the active ones; the vacuum expectation value of $\varphi$ gives the sterile neutrinos Majorana masses, $M = \sqrt{2} v_\varphi Y_N$. We compute the one-loop self-energy of the sterile states, which gives \equ{meff_model2}, and the running of the parameters, which shows that the contributions of the sterile sector to the running of the mixing cancel.

In the basis where $Y_N = {\rm diag}(y_k)$, we write each sterile state as a four-component Majorana spinor, $N_k = (n_k, \bar n_k)^T$, built from a two-component spinor, $n_k$, and its conjugate. With $\varphi = (v_\varphi + \sigma + ia)/\sqrt{2}$, \equ{model2_lagrangian} couples the sterile states to $\sigma$ and $a$ through $-\frac{1}{2}\lambda_k \sigma \overline{N_k} N_k - \frac{1}{2} i\lambda_k a \overline{N_k}\gamma_5 N_k$, with $\lambda_k = \sqrt{2} y_k$. The one-loop self-energy of $N_k$ is
\begin{equation}
 \Sigma_k(p) = -\frac{\lambda_k^2}{16\pi^2} \left[ \slashed{p} \left( B_1^\sigma + B_1^a \right) + M_k \left( B_0^\sigma - B_0^a \right) \right] ,
\end{equation}
where $B_{0,1}^X \equiv B_{0,1}(p^2; M_k, m_X)$, with $X = \sigma, a$, are the standard Passarino--Veltman loop functions~\cite{Passarino:1978jh}. The chirality-flipping part is ultraviolet-finite, since the divergences cancel between $\sigma$ and $a$. For the light states, at $p^2 \simeq 0$, only that part enters, through $M_k \to M_k^{\rm eff}$ in the seesaw formula; the kinetic part multiplies $\slashed{p}$ and contributes only at $\mathcal{O}(m_D^2/M^2)$. With $B_0(0; M, m) = \Delta_{\rm UV} + 1 - [M^2\ln(M^2/\mu^2) - m^2\ln(m^2/\mu^2)]/(M^2 - m^2)$ and $m_a = 0$, where $\Delta_{\rm UV}$ contains the ultraviolet divergence, this gives \equ{meff_model2}.

As in Sec.~\ref{sec:examples-model2}, the divergent parts of the self-energies make the normalization of each field run with $\mu$; the rate of this running is the anomalous dimension of the field~\cite{Machacek:1983tz, Luo:2002ti}. For the sterile states, it comes from the term of $\Sigma_k$ proportional to $\slashed{p}$; for $\varphi$, from its own self-energy, a loop of sterile states. In the basis $Y_N = {\rm diag}(y_k)$,
\begin{equation}
 \gamma_{n_k} = \frac{2 y_k^2}{16\pi^2} \,, \qquad
 \gamma_\varphi = \frac{2\sum_j y_j^2}{16\pi^2} \,.
 \label{equ:model2_gammas}
\end{equation}
Here, $\gamma_{n_k}$ and $\gamma_\varphi$ are the anomalous dimensions of the sterile field $n_k$ and of the scalar, respectively; $\gamma_{v_\varphi} \equiv d\ln v_\varphi/d\ln\mu$ describes the running of the vacuum expectation value of the scalar. Only $\gamma_{n_k}$ depends on the index $k$; $\gamma_\varphi$ and $\gamma_{v_\varphi}$ are the same for all $k$. We define $\gamma_{\rm univ} \equiv \gamma_\varphi + \gamma_{v_\varphi}$.

Since the one-loop $\varphi N N$ vertex has no divergence, $y_k$ runs only through the normalizations of the fields in the Yukawa term, two sterile fields and $\varphi$~\cite{Machacek:1983fi, Luo:2002ti}:
\begin{eqnarray}
 \frac{dy_k}{d\ln\mu} &=& y_k \left( 2\gamma_{n_k} + \gamma_\varphi \right) \nonumber \\
 &=& \frac{1}{16\pi^2} \left( 4 y_k^3 + 2 y_k \sum_j y_j^2 \right) .
\end{eqnarray}
This reproduces \equ{beta_yn} coefficient by coefficient, which confirms the $\beta$ function of \Refe~\cite{Babu:2021cxe}.

The same rule applies to the masses in the seesaw formula: each runs with the anomalous dimensions of the fields it multiplies. The Dirac mass term, $m_D \bar\nu N$, contains one active and one sterile field, so
\begin{equation}
 \frac{d(m_D)_{\alpha k}}{d\ln\mu}
 =
 (\gamma_L)_{\alpha\alpha'} (m_D)_{\alpha' k} + (m_D)_{\alpha k} \, \gamma_{n_k} \,,
 \label{equ:model2_betamD}
\end{equation}
where $\gamma_L$ is the anomalous dimension of the active lepton doublet. It comes from loops with the Higgs and electroweak bosons, so it vanishes below the electroweak scale, where the terrestrial analysis of \Refe~\cite{Babu:2021cxe} applies. There, $m_D$ still runs, through $\gamma_{n_k}$, because $\varphi$ is light and not integrated out. The Majorana mass, $M_k = \sqrt{2} v_\varphi y_k$, multiplies two sterile fields and runs with $v_\varphi$:
\begin{equation}
 \frac{d\ln M_k}{d\ln\mu}
 =
 2\gamma_{n_k} + \gamma_\varphi + \gamma_{v_\varphi}
 =
 2\gamma_{n_k} + \gamma_{\rm univ} \,.
 \label{equ:model2_betaM}
\end{equation}

The seesaw mass matrix is a sum of terms, one for each sterile state, $M_\nu = \sum_k T_k$, with
\begin{equation}
 (T_k)_{\alpha\beta} \equiv -\frac{(m_D)_{\alpha k} (m_D)_{\beta k}}{M_k} \,.
\end{equation}
Each term contains the field $n_k$ twice in its numerator, through the two Dirac masses, and twice in its denominator, through $M_k$. By \equ{model2_betamD}, the two Dirac masses make $T_k$ run at the rate $2\gamma_{n_k}$; by \equ{model2_betaM}, the Majorana mass in the denominator makes it run at the rate $-(2\gamma_{n_k} + \gamma_{\rm univ})$. Hence, the normalization of $n_k$ cancels:
\begin{equation}
 \frac{dT_k}{d\ln\mu}
 =
 \left[ 2\gamma_{n_k} - \left( 2\gamma_{n_k} + \gamma_{\rm univ} \right) \right] T_k
 =
 -\gamma_{\rm univ} \, T_k \,,
 \label{equ:model2_cancel}
\end{equation}
up to the $\gamma_L$ terms. Summing over $k$,
\begin{equation}
 \frac{dM_\nu}{d\ln\mu}
 =
 \gamma_L M_\nu + M_\nu \gamma_L^T - \gamma_{\rm univ} M_\nu \,.
 \label{equ:model2_total}
\end{equation}
The $\gamma_L$ terms are the standard running of Sec.~\ref{sec:general-running}, unrelated to $Y_N$. The last term multiplies all elements of $M_\nu$ by the same factor, which leaves its eigenvectors unchanged. Therefore, below the electroweak scale, where $\gamma_L = 0$, the mixing does not run.

Reference~\cite{Babu:2021cxe} instead holds $Y_\nu$, and hence $m_D$, fixed. It runs $Y_N$ but not $v_\varphi$. Therefore, its Dirac masses do not run; its Majorana masses run with $\gamma_\varphi$ in place of $\gamma_{\rm univ}$. Below the electroweak scale, its running of $M_\nu$ differs from \equ{model2_total} by
\begin{eqnarray}
 \lefteqn{\left. \frac{dM_\nu}{d\ln\mu} \right|_{\text{\Refe~\cite{Babu:2021cxe}}}
 - \left. \frac{dM_\nu}{d\ln\mu} \right|_{\rm consistent}} \nonumber \\
 \qquad &=&
 \gamma_{v_\varphi} M_\nu - 2 \sum_k \gamma_{n_k} T_k \nonumber \\
 &=&
 \gamma_{v_\varphi} M_\nu - \frac{1}{4\pi^2} \sum_k y_k^2 \, T_k \,.
 \label{equ:model2_leftover}
\end{eqnarray}
The first term multiplies $M_\nu$ by a common factor and does not change the mixing. The second weights each $T_k$ by $y_k^2$, so it changes their relative sizes and rotates the eigenvectors; it produces the running of the mixing shown in \figu{running_examples}C for $Y_\nu$ held fixed. For benchmark BP2, with $Y_N(Q_p) = {\rm diag}(0.2, 0.5, 0.7)$, it comes mainly from the second and third sterile states, with $y_k^2 = 0.25$ and $0.49$. Letting the Dirac masses run, as \equ{model2_betamD} requires, removes this term exactly.

\section{Numerical inputs}
\label{app:numerics}

Unless stated otherwise, we use the NuFIT~6.1 best-fit values for normal ordering, from the fit that includes Super-Kamiokande atmospheric data~\cite{Esteban:2024eli, NuFIT61}: $\sin^2\theta_{12} = 0.3088$, $\sin^2\theta_{23} = 0.470$, $\sin^2\theta_{13} = 0.02248$, $\delta_{\rm CP} = 212^\circ$, $\Delta m_{21}^2 = 7.537 \times 10^{-5}$~eV$^2$, and $\Delta m_{31}^2 = 2.511 \times 10^{-3}$~eV$^2$. We diagonalize complex symmetric mass matrices by Takagi factorization, $U^T m U = {\rm diag}(m_i)$, with $m_i \geq 0$.

For the MSSM example, we use vanishing Majorana phases. For the HNL example, we use $v = 246.22$~GeV, $M_Z = 91.1876$~GeV, and $M_h = 125.25$~GeV.

For Model~2, we use the inputs and conventions with which \Refe~\cite{Babu:2021cxe} computed its results for Model~1: $\sin^2\theta_{12} = 0.310$, $\sin^2\theta_{13} = 0.022$, $\Delta m_{21}^2 = 7.53 \times 10^{-5}$~eV$^2$, $m_1 = 0.05$~eV, and $Y_N(Q_p) = {\rm diag}(0.2, 0.5, 0.7)$. We build $Y_\nu$ from the parametrization $Y_\nu \propto U \sqrt{m}\, R\, Y_N^{-x/2}$. Here, $\sqrt{m} \equiv {\rm diag}(\sqrt{m_1}, \sqrt{m_2}, \sqrt{m_3})$; $x = -1$ for Model~2 and $x = +1$ for Model~1; $R = R_{13}(\xi_3) R_{23}(\xi_2) R_{12}(\xi_1)$ is an orthogonal matrix, with $R_{ij}(\xi)$ a rotation by the angle $\xi$ in the $ij$ plane; and $U$ includes the Majorana phases through the factor ${\rm diag}(1, e^{i\tilde\alpha}, e^{i\tilde\beta})$. Reference~\cite{Babu:2021cxe} writes this relation with $U^\dagger$ in place of $U$, but its convention for the mixing matrix, $U^\dagger M_\nu U^\ast = {\rm diag}(m_i)$, requires $U$; with $U$, we reproduce its Fig.~8. Its benchmark BP2 has $(\delta_{\rm CP}, \tilde\alpha, \tilde\beta, \xi_1, \xi_2, \xi_3, \theta_{23}) = (1.18, 0.24, 1.64, 5.48, 2.076, 1.85, 0.86)$~rad and $\Delta m_{31}^2 = 2.525 \times 10^{-3}$~eV$^2$.

\bibliography{refs}

\end{document}